\documentclass[aps,prfluids,superscriptaddress]{revtex4-2}

\usepackage{amsmath}
\usepackage{amssymb}
\usepackage{graphics}
\usepackage{graphicx}
\usepackage{color}
\usepackage{tikz}
\usetikzlibrary{shapes}
\usepackage{hyperref}
\usepackage{CJKutf8}

\newcommand{\q}[1]{‘#1’}
\newcommand{\upchi}{\raisebox{2pt}{$\chi$}}
\newcommand{\Rey}{\text{Re}}
\newcommand{\Ga}{\text{Ga}}
\newcommand{\Bo}{\text{Bo}}
\newcommand{\thickline}[1]{\protect\tikz[baseline=-0.5ex]{\protect\draw[line width=1mm, draw= #1] (0,0) -- (0.5cm,0);}}
\newcommand{\full}[1]{\protect\tikz[baseline=-0.5ex]{\protect\draw[line width=0.5mm, draw= #1] (0,0) -- (0.5cm,0);}}
\newcommand{\dashed}[1]{\protect\tikz[baseline=-0.5ex]{\protect\draw[line width=0.5mm, draw= #1, dashed] (0,0) -- (0.5cm,0);}}
\definecolor{DEPART}{HTML}{00BFFF}
\definecolor{BALANCE}{HTML}{41DC8E}
\definecolor{BOUNCE}{HTML}{FF3535}

\newcommand{\bitb}{\protect\tikz[baseline=-0.5ex]{\protect\node[regular polygon, thick,regular polygon sides=5,draw=black,scale =0.6] at (0,0){};}}

\newcommand{\bilb}{\protect\tikz[baseline=-0.5ex]{\protect\node[circle,thick,draw=black,scale=0.7] at (0,0){}; \protect\draw[fill=black,scale=0.7,draw=black] (0,0)-- (0:4pt) arc (0:180:4pt) -- cycle;}}

\begin{document}


\title{Path of a pair of deformable bubbles rising initially in line and close to a vertical wall}


\author{Haochen Huang \begin{CJK*}{UTF8}{gbsn}(黄\end{CJK*}\begin{CJK*}{UTF8}{bsmi}澔辰)\end{CJK*}}
\affiliation{State Key Laboratory for Strength and Vibration of Mechanical Structures, School of Aerospace, Xi’an Jiaotong University (XJTU), Xi’an 710049, China}
\author{Pengyu Shi}
\email{p.shi@hzdr.de}
\affiliation{Institut de M\'ecanique des Fluides de Toulouse (IMFT), Universit\'e de Toulouse, CNRS, F-31400 Toulouse, France}
\affiliation{Helmholtz-Zentrum Dresden-Rossendorf (HZDR), Institute of Fluid Dynamics, 01328 Dresden, Germany}
\author{Nina Elkina}
\affiliation{Helmholtz-Zentrum Dresden-Rossendorf (HZDR), Department of Information Services and Computing, 01328 Dresden, Germany}
\author{Henrik Schulz}
\affiliation{Helmholtz-Zentrum Dresden-Rossendorf (HZDR), Department of Information Services and Computing, 01328 Dresden, Germany}
\author{Jie Zhang\begin{CJK*}{UTF8}{gbsn}(张杰)\end{CJK*}}
\email{j\_zhang@xjtu.edu.cn}
\affiliation{State Key Laboratory for Strength and Vibration of Mechanical Structures, School of Aerospace, Xi’an Jiaotong University (XJTU), Xi’an 710049, China}


\date{\today}

\begin{abstract}
It is known that in an unbounded fluid, the inline configuration of a freely rising bubble pair is often unstable with respect to lateral disturbances.
This work numerically examines the stability of this configuration in the presence of a nearby vertical wall. The focus is on moderately inertial regimes, where two bubbles rising initially in line typically separate laterally from each other under unbounded conditions. In the presence of the wall, our results indicate that while the path of the bubble pair predominantly separates laterally, the plane of separation largely depends on the wall-bubble interaction. This interaction involves a competition between two distinct effects, with the dominance determined by the ratios of buoyancy-to-viscous and buoyancy-to-capillary forces, which define the Galilei $(\Ga)$ and Bond $(\Bo)$ numbers, respectively. When $\Bo$ is below a critical $\Ga$-dependent threshold, irrotational effects dominate, initially stabilizing both bubbles near the wall until horizontal separation among them occurs in the wall-parallel plane. Conversely, at higher $\Bo$, vortical effects dominate such that both bubbles migrate away from the wall. During the departure, asymmetric interactions cause the wall-normal velocities of the two bubbles to differ, leading to horizontal separation in the wall-normal plane. These two separation motions, both newly identified in the present study, are found to result from two distinct mechanisms: one associated with the shear flow generated in the gap separating the wall and the leading bubble, which attracts the trailing bubble toward the wall, and the other linked to vortex shedding from the leading bubble, which promotes the trailing bubble’s faster escape from the wall. A slight angular deviation favours separation in the wall-parallel plane, promoting the formation of a near-wall, bubble-rich layer as observed in prior investigations of buoyancy-driven, bubble-laden flows.
\end{abstract}


\maketitle

\section{Introduction}\label{sec:1intro}
Wall-bounded bubble-laden flows are encountered in many technical processes, such as mixing in chemical reactors, cooling of nuclear reactors, and aeration for water treatment. Owing to the presence of the wall, the bubble distribution is usually inhomogeneous, particularly in bubble columns or upward channel flows, where bubbles tend to accumulate towards the wall \citep{1975_Serizawa, 1987_Wang, 1999_Riviere, 2006_Lu, 2008_Takagi, 2013_Lu, 2015_Santarelli, 2019_du}. Within the bubble-rich, near-wall layer, bubbles might cluster either horizontally or vertically, depending on the bubble size and the level of contamination \citep{2001_Zenit, 2003_Bunner, 2011_Takagi, 2013_Lu, 2021_Maeda}. Such near-wall clusters can extend several tens of bubble radii and may therefore alter the flow structures across a wide range of length scales \citep{2018_Risso, 2022_Ma}. Our current understanding of the mechanisms responsible for these inhomogeneous near-wall bubble distributions remains limited. In particular, the entire problem involves interactions between bubbles, liquid, and the wall, and cannot be fully understood by considering some idealized, isolated conditions \citep{2008_Yin, 2011_Takagi, 2013_Lu}.\par 

At the local scale, the primary step in the understanding of these interactions consists in considering the rise of a bubble pair in a quiescent fluid partially bounded by a flat wall.
Prior relevant work can be divided into two categories, each considering the same problem but with a certain level of simplifications. The first series of works focuses on bubble pair interactions in an unbounded fluid, aiming to elucidate the clustering mechanisms. For an overview of the extensive prior work on this problem, we refer the readers to \citet{2021_zhang, 2022_Zhang}. One of the key outcomes from these studies is the identification of a drafting–kissing–tumbling mechanism. 
This mechanism drives a pair of nearly spherical bubbles, rising at $\Rey=O(10)$ (with $\Rey$ the Reynolds number based on the bubble diameter $2R$ and the slip velocity $u$), to eventually align side by side, promoting horizontal clustering of bubbles in the bulk region of the channel \citep{2003_Bunner, 2013_Lu}.
Additionally, a distinct mechanism stems from the reversal of the shear-induced lift when the amount of vorticity generated at the bubble surface is large.
This lift-reversal mechanism, notable for significantly oblate bubbles \citep{2009_Adoua} or suspensions in the presence of surfactants \citep{2008_Fukuta}, stabilizes the inline configuration of a bubble pair and thus promotes the formation of vertical clusters \citep{2003_Bunner, 2010_Mercado, 2013_Tagawa, 2023_Atasi}.\par

The second stream of research considers the near-wall rising motion of a single bubble, which provides insights into the mechanisms driving near-wall bubble accumulation. We refer the readers to \citet{2003_Takemura} and, more recently, \citet{2024_Shi_b} for an overview of this body of research. In moderately inertial regimes where the rising path remains rectilinear under unbounded conditions \citep{2023_Bonnefis_b}, the wall-bubble interaction is governed by the competition between irrotational and vortical effects. For weakly deformed bubbles rising at $\Rey=O(100)$, the irrotational effects dominate \citep{1976_Wijngaarden, 
1993_Kok_a, 2003_Legendre}, stabilizing the bubble at a (mean) wall-normal position close to the wall \citep{2024_Shi_a}. This stabilization is indeed one of the key causes of the near-wall void peaking observed in bubble columns and air-lift reactors \citep{2016_Ziegenhein, 2018_Besagni}. However, the path of these near-wall rising bubbles may not always be rectilinear. Owing to history effects associated with wall-induced vortex shedding, the bubble may oscillate laterally with a displacement amplitude of $O(R)$ \citep{2001_Vries, 2002_Hosokawa, 2003_Takemura, 2024_Shi_b}, where $R$ is the bubble radius.\par

Albeit the many progresses achieved in each stream of investigations, a thorough physical understanding of the near-wall bubble distribution is still challenging. Note that such knowledge cannot be obtained by merely synthesizing the outcomes of these prior works. Consider, for instance, the pair interaction in the presence of a wall: the bubbles may oscillate laterally in the wall-normal plane, meaning their path evolutions could differ significantly from those in unbounded conditions. This also applies to wall-bubble interactions: the hydrodynamic field around a near-wall rising bubble is altered by the presence of its neighbours, which changes the dynamics of competition between the irrotational and the vortical effects. The present work aims to provide a detailed understanding of these interaction processes. For this purpose, we examine the problem of a pair of deformable bubbles rising initially inline in the vicinity of a vertical wall. We carried out high-resolution, three-dimensional, time-dependent computations that allow for a complete interplay of pair and wall-bubble interactions. The main questions to be addressed are:

(i): What are the typical path evolutions and, compared with prior knowledge in an unbounded fluid, what is the role of the wall?

(ii): What information can be inferred from the final geometry of the bubble pair in understanding the inhomogeneous near-wall bubble distributions?

The paper is organized as follows: in \S~\ref{sec:2state}, we formulate the problem, outline the governing parameters, and briefly introduce the numerical approach applied. Section \ref{sec:3res} is divided into five subsections: the first three detail the numerical results obtained, followed by a discussion in \S~\ref{sec:3-4par_eff} on the effects of initial angular deviations. Section \ref{sec:3-5mac} addresses the potential implications for inhomogeneous near-wall bubble distributions. Concluding remarks are given in \S~\ref{sec:4conclude}.

\section{Statement of the problem and outline of the numerical approach}\label{sec:2state}

\subsection{Problem statement}
\begin{figure}
    \centering
    \includegraphics[height=7.5cm]{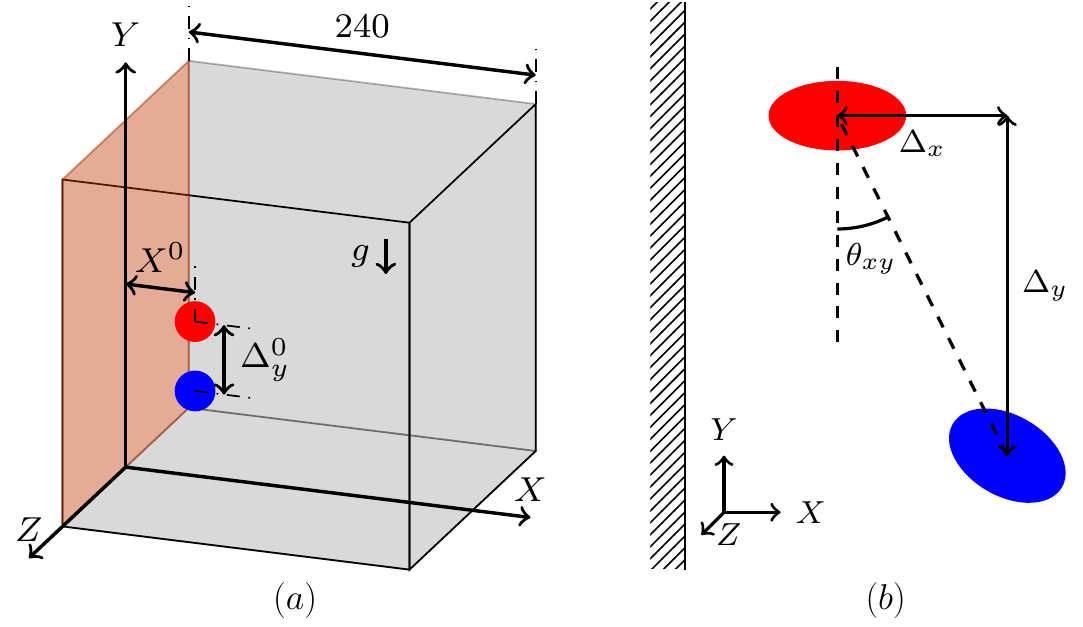}
    \caption{Sketch of the problem.}
    \label{fig:1sketch}
\end{figure}

We consider the buoyancy-driven motion of a pair of initially spherical bubbles with radius $R$, released inline and close to a vertical wall in an otherwise quiescent liquid. For brevity, throughout the rest of this paper, we refer to the leading bubble, i.e., the one initially released at a higher vertical level, as \q{LB}, and the trailing bubble, which initially rises after the LB, as \q{TB}. The velocities of the two bubbles are distinguished by superscripts, namely $\mathbf{u}^\text{LB}$ and $\mathbf{u}^\text{TB}$ for the LB and the TB, respectively. We define the Cartesian frame of reference $(X, Y, Z)$, as illustrated in figure \ref{fig:1sketch}, with the wall located at $X=0$, the gravitational acceleration $\mathbf{g}$ aligned along $-Y$, and both bubbles released within the wall-normal plane at $Z=0$ along the vertical line $X=X^0$. The initial vertical separation between the centres of the two bubbles is $\Delta_y^0$. A superscript 0 is used to emphasize that these are the values at $t=0$.

Starting from rest, the two bubbles begin to rise freely under the influence of buoyancy. Given the disturbances that arise in the wall-normal [i.e., the $(X, Y)$] and the wall-parallel [i.e., the $(Y, Z)$] planes, the path of the two bubbles may become three-dimensional during their ascent. To characterize their motion, we denote by $\Delta_x$ and $\Delta_z$ the separation distances between the bubble centres along the $X$ and $Z$ axes, respectively. Here, positive values of $\Delta_x$ and $\Delta_z$ indicate that the LB is located at a larger coordinate value along that respective dimension. To better describe the relative orientation of the two bubbles, we also introduce two inclination angles (see, e.g., figure \ref{fig:1sketch}$b$): $\theta_{xy}=\tan^{-1}(\Delta_x/\Delta_y)$ and $\theta_{zy}=\tan^{-1}(\Delta_z/\Delta_y)$. Both angles would be zero if the two bubbles are inline.

The liquid and bubble motions are described by the incompressible one-fluid Navier–Stokes equations. Throughout this work, we have set the density and (dynamic) viscosity ratios of the liquid to the gas phase at 1000 and 100, respectively, to simulate a gas bubble rising in a liquid. At the initial state, the wall-normal separation and the vertical distance between the two bubbles are set at $X^0 = 1.5R$ and $\Delta_y^0 = 8R$, respectively (see figure \ref{fig:1sketch}$a$). 
These two geometrical parameters are selected to facilitate comparison with results from our prior work, where the same parameters were used to predict the motion of an unbounded bubble pair \citep{2021_zhang} and a single near-wall rising bubble \citep{2024_Shi_b}. The influences of $X^0$ and $\Delta_y^0$ on the path of the bubble pair are examined in \S~\ref{sec:3-4par_eff} and Appendix \ref{sec:appA2}.

Based on these specifications, the problem is governed by the Galilei number $\Ga$ and the Bond number $\Bo$, which are defined as
\begin{equation}
\Ga=\frac{\rho_l g^{1/2} R^{3/2}}{\mu_l}, \quad \Bo=\frac{\rho_l g R^2}{\gamma},
\end{equation}
where $\rho_l$ and $\mu_l$ represent the density and dynamic viscosity of the liquid, respectively, $g$ is the magnitude of gravitational acceleration, and $\gamma$ is the interfacial tension. To establish a connection with prior theoretical research, we will also frequently characterize a given case by the Reynolds number and the aspect ratio determined by considering the near-wall rising motion of only the LB.
These two parameters are defined as $\Rey=\rho_l u_T (2R)/\mu_l$ and $\upchi=b/a$, where $u_T$ is the (mean) terminal rising velocity, and $b$ and $a$ are the major and minor axes of the bubble, respectively. We focus on the parameter range of $0.02 \leq \Bo \leq 1$ and $10 \leq \Ga \leq 30$. The corresponding ranges of the Reynolds number and aspect ratio covered are $25 \lesssim \Rey \lesssim 200$ and $1.01 \lesssim \upchi \lesssim 2.1$. 

\subsection{Numerical approach}
The open-source solver \emph{Basilisk} \citep{2009_Popinet, 2015_Popinet} has been utilized to solve the flow around the two bubbles. To fully capture the three-dimensional effects mentioned previously, a full three-dimensional domain has been employed, as illustrated in figure \ref{fig:1sketch}$(a)$. The computational domain is a cubic box with a size of $240R$. Following \citet{2021_zhang}, the two bubbles are released close to the left bottom corner, with the initial separation between the TB and the bottom surface set at $6R$. A no-slip boundary condition has been imposed on all surfaces of the domain; the resulting confinement effects have been confirmed to be negligibly small \citep{2024_Shi_b}. The Volume-of-Fluid (VOF) method, along with the continuum-surface-force algorithm \citep{2018_popinet}, has been employed to capture the gas-liquid interface and compute the capillary force. Specifically, the presence of the bubbles is characterized by the VOF function $C(\mathbf{X}, t)$, with $C=1$ inside the bubble and $C=0$ in the liquid. The local density and viscosity are approximated using their respective arithmetic and harmonic means, defined as $\rho=C \rho_g + (1-C)\rho_l$ and $1/\mu=C/\mu_g + (1-C)/\mu_l$, with the subscript $g$ denoting the gas phase. Lastly, the rigid wall is assumed to be fully hydrophilic, thereby preventing the formation of a three-phase contact line. Numerically, this is facilitated by imposing $C=0$ on $X=0$ \citep{2024_Shi_b}, ensuring that there is at least one grid cell with $C<1$ between the gas phase and the wall.

Some details about the grid resolution are provided below. The grid is dynamically refined using an adaptive mesh refinement (AMR) technique \citep{2018_Hooft}. The reference parameters chosen for the refinement are the liquid velocity $\boldsymbol{U}$ and the VOF function $C$, with the respective thresholds for refinement set at $10^{-2}$ and $10^{-3}$. The minimum and maximum grid sizes are $\Delta_{min} \approx R/68$ and $\Delta_{max} \approx 4R$, respectively. The grid resolution, as per the above settings, is found to be sufficient in capturing the wall-bubble interaction \citep{2024_Shi_b}. To ensure that the pair interaction between the two bubbles is also reasonably resolved when they are sufficiently close, we employ a topology-based AMR strategy \citep{2019_zhang}. According to this strategy, when two bubbles are close enough that the thickness of the thin film separating them is smaller than $\Delta_{min}$, the grid within the film is further refined to $\delta_{min} = \Delta_{min}/2^2 \approx R/272$. The two bubbles will coalesce only if the gap separating them is smaller than $\delta_{min}$. This refinement strategy has been shown to be effective in suppressing numerical coalescence and resolving the initial stages of drainage, as demonstrated in \citet{2021_zhang}, where two bubbles initially rising inline in an unbounded domain was considered. For a graphical illustration of the grid structures, readers are referred to \citet[appendix A therein]{2024_Shi_b} for the grid structure in the vicinity of the wall and to \citet[Fig. 3 therein]{2021_zhang} for the grid refinement in the thin film when two bubbles are close to coalescence.

The reliability of the numerical approach in accurately capturing the wall-bubble and pair interactions has been confirmed in our prior work \citep{2021_zhang, 2024_Shi_b}. For the wall-bubble interaction, numerical simulations have been carried out considering a single bubble rising close to a vertical or inclined rigid wall. The obtained results, including the type of near-wall rising motion, the mean rising velocity, and the aspect ratio, are in good agreement with prior experimental data \citep{2003_Takemura, 2014_Kosior, 2016_Barbosa}. Regarding the pair interaction, we considered the problem of two nearly spherical bubbles ($\Bo=0.005$) rising inline. The results for the final equilibrium distance align well with prior numerical results from \citet{1994_Yuan} and \citet{2011_Hallez}. The sufficiency of the numerical approach, as confirmed in these prior works, permits the use of the same numerical method in the current investigation to examine the combined problem where wall-bubble and pair interactions occur simultaneously.

The simulations for the cases with $\Bo \geq 0.2$ were carried out on the HPC cluster at XJTU MFM lab, executed by two 32-core Intel\textsuperscript{\textregistered} Xeon\textsuperscript{\textregistered} Platinum 8370C CPU processors running at 2.80 GHz. A single case run covering a physical time of $80(R/g)^{1/2}$ takes approximately 30 days (e.g., $(\Ga, \Bo) = (15, 0.2)$). 
The remaining cases where $\Bo < 0.2$ were carried out on the \emph{Hemera} HPC cluster at HZDR. For details and computational costs of \emph{Hemera}, we refer readers to \cite{2024_Shi_b}, which used the same HPC cluster to conduct simulations on the near-wall rising motion of a single bubble within the same $(\Ga, \Bo)$ regime as this work.
Note that, unlike in this previous work where the flow remained symmetric with respect to the wall-normal plane (allowing for considering only half of the computational domain), such symmetry usually does not persist in the present work. For all simulations carried out in the present work, a full computational domain is employed. Due to this difference, the typical number of grid cells at a given $(\Ga, \Bo)$ is usually 3 to 4 times larger than in that previous work. For instance, at $(\Ga, \Bo) = (25, 0.05)$, the number of grid cells is about 12 million when the wake of the two bubbles is fully developed, compared to about 3.3 million in that previous work. Consequently, at a given $(\Ga, \Bo)$, the wall time consumed is approximately twice that of the previous work.

In the following sections, we extensively employ dimensionless flow parameters to describe the motion of the bubble. The characteristic scales for length, velocity, time, and vorticity used in the normalization are $R$, $(gR)^{1/2}$, $(R/g)^{1/2}$, and $(g/R)^{1/2}$, respectively.

\section{Results and discussion}\label{sec:3res}
\subsection{Overview of the results}\label{sec:3-1ove}
The fate of a bubble pair initially rising inline is well understood in unbounded conditions \citep{2021_zhang}. In this context, the inline configuration is stable only beyond a critical $\Ga$-dependent Bond number, $\Bo_{c1}(\Ga)$, where the two bubbles coalesce through a head-on approach (i.e., without side escape). For $\Bo < \Bo_{c1}(\Ga)$, the inline configuration breaks into two distinct asymmetric configurations depending on the $\Ga$. For $\Ga \lesssim 12$, the bubble pair follows a drafting–kissing–tumbling scenario (hereafter abbreviated as DKT scenario), which eventually yields a planar side-by-side motion. For larger $\Ga$, the bubble pair evolves in such a way that the TB escapes laterally from the wake of the LB without significantly altering the path of the latter. These prior results serve as a good reference to examine the effects of the wall on the fate of a bubble pair, which will be outlined in the following.

\begin{figure}
    \centering
    \includegraphics[height=8cm]{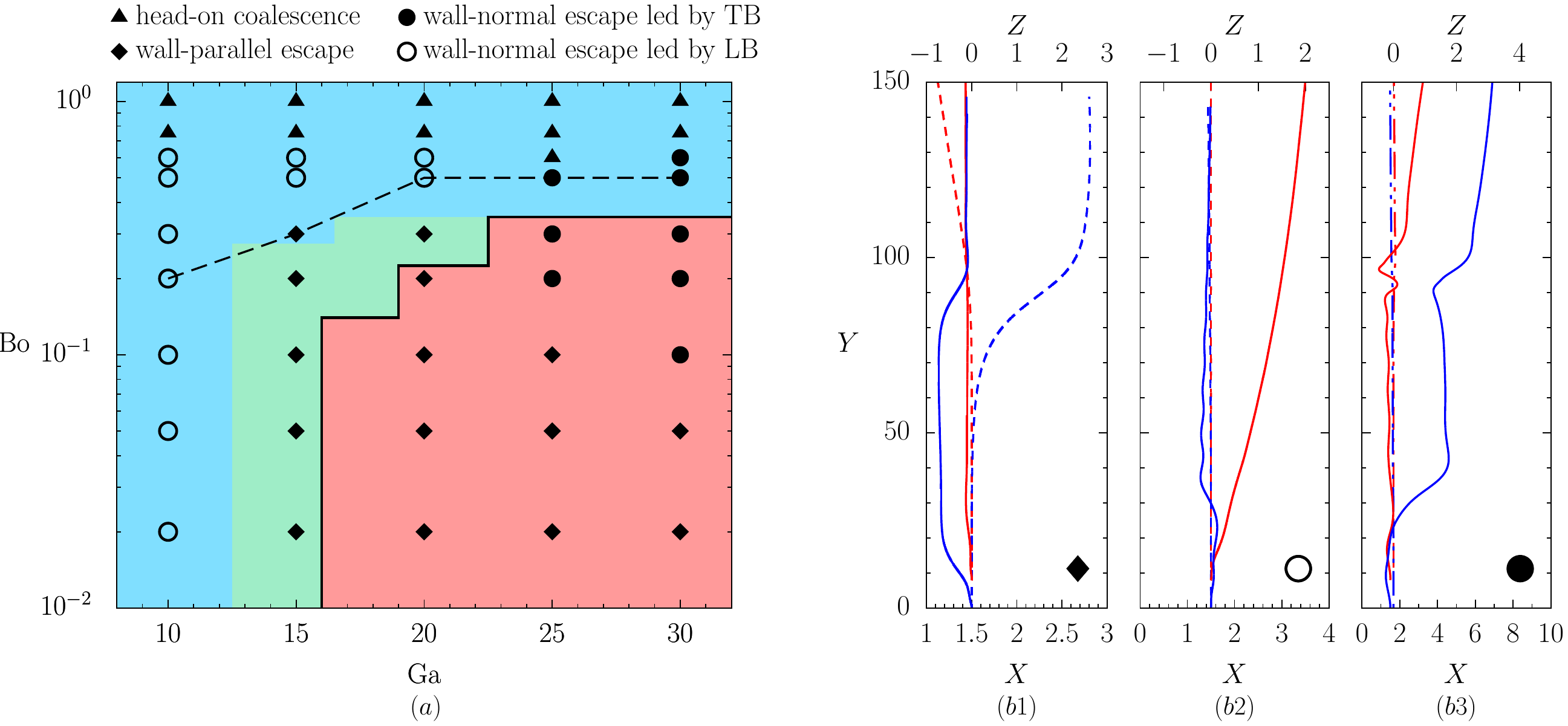}
   \caption{Different types of pair motion observed in the simulations. $(a)$ Phase diagram in the $(\Ga, \Bo)$ plane. $(b1-b3)$ Path of the bubble pair in three distinct scenarios, namely, wall-parallel escape $(b1)$, wall-normal escape led by the LB $(b2)$, and by the TB $(b3)$. In $(a)$, different symbols correspond to different types of motion in the pair configuration, whereas different background colours denote the motions in the corresponding single-bubble configuration. Specifically,  
   \thickline{DEPART}: migration away from the wall; \thickline{BALANCE}: rest close to the wall; \thickline{BOUNCE}: periodic near-wall bouncing. Also in $(a)$: the dashed black line denotes the critical Bond number, $\Bo_{c1}(\Ga)$, beyond which head-on coalescence occurs in the pair configuration without the presence of the wall; the solid black line separates the stable and unstable regimes; in the latter regime, vortex shedding occurs.} In $(b1-b3)$, the three cases from left to right correspond to $(\Ga,\Bo)=(15,0.2)$, $(15,0.5)$, and $(25,0.3)$. In each of these three panels, solid (respectively, dashed) lines denote the path in the $(X,~Y)$ [respectively, $(Z,~Y)$] plane, with the tick labels for $X$ (respectively, $Z$) appearing at the horizontal bottom (respectively, top) axis. Lines in red and blue denote the path of the LB and TB, respectively.
    \label{fig:2phase}
\end{figure}

Results for the type of motions obtained in the wall-bounded configuration are illustrated in figure \ref{fig:2phase} in the $(\Ga, \Bo)$ phase diagram. In the high-$\Bo$ regime, the head-on coalescence scenario still manifests, except that now the critical Bond number at a given $\Ga$ is larger. Specifically, $\Bo_{c1}(\Ga)$ is about 0.7 and almost independent of $\Ga$ in the wall-bounded configuration, compared with only about 0.2 and 0.5 at $\Ga=10$ and 30, respectively, in the corresponding unbounded condition. For $\Bo<\Bo_{c1}(\Ga)$, the bubble pair always separates horizontally through an \q{escape} approach, even for $\Ga \lesssim 12$ where they follow a DKT scenario in the unbounded condition. These changes in the type of motion are similar to those induced by an initial angular deviation as assessed in the unbounded condition in \citet[see section 7.3 therein]{2021_zhang}. There, with an initial angular deviation, the coalescence can be avoided up to a larger Bond number, and the DKT scenario can be entirely prevented. This similarity may be understood by noting that the presence of the wall makes the wake of the LB strongly asymmetric. Given this wall-induced asymmetry, the sideways force experienced by the two bubbles starts to differ once the wake of the LB has been advected downstream over a distance of $O(X^0)$.

To better illustrate the effects of the wall-induced asymmetry, let us first overview the path of a single bubble rising close to the wall. In this context, the motion is governed by the interplay between the vortical effect, which leads to a repulsive lateral force, and the irrotational one, which generates an attractive lateral force. The vortical effect dominates for $\Bo$ beyond a critical, $\Ga$-dependent value, $\Bo_{c2}(\Ga)$, such that all bubbles in this high $\Bo$ regime migrate away from the wall \citep{2002_Takemura, 2003_Takemura, 2024_Estepa, 2024_Shi_b}. For $\Bo < \Bo_{c2}(\Ga)$, where the two mechanisms contribute comparably, the bubble maintains a close distance to the wall while ascending, following either a vertical path or a laterally oscillating path. In the latter case, the flow is unsteady, characterized by strong vortex shedding.

Now let us have a closer look at the effects of the wall-induced asymmetry. Given that the propagation towards the wall of the disturbance is inhibited, the wake of the LB on the wall-bounded side is elongated (see the vortical structure in, e.g., figure \ref{fig:4wpe_r_velo}$a$). When this asymmetric wake is advected downwards to the TB, it leads to an attractive shear-induced lift on the TB. Taken together with the picture of interaction in the single-bubble condition, this attractive contribution cooperates with the irrotational effect, stabilizing the TB to a (mean) position closer to the wall. Consequently, the escape of the TB in the wall-normal plane is strongly inhibited, and the horizontal separation of the two bubbles is more favoured in the wall-parallel plane. This is indeed the scenario seen in most of the cases where the bubble stabilizes close to the wall in the corresponding single-bubble configuration (see, e.g., figure \ref{fig:2phase}$(b1)$ for the corresponding path). On the other hand, when the vortical effect is strong enough such that the bubble departs from the wall in the corresponding single-bubble configuration, the separation takes place in the wall-normal plane, owing to the faster departing velocity of the LB (see, e.g., figure \ref{fig:2phase}$(b2)$ for the corresponding path). Finally, in both situations, there is an exceptional scenario where the separation takes place in the wall-normal plane, as in the second situation, but the separation is led by the TB, not the LB (see, e.g., figure \ref{fig:2phase}$(b3)$ for the corresponding path). This third scenario occurs at large $\Ga$ and moderate $\Bo$ (see the solid cycle symbols in figure \ref{fig:2phase}$a$). In this regime, the TB gains a finite departing velocity prior to the pair interaction, enabling it to overcome the wall-ward shear-induced lift and reach the unbounded side of the LB's wake. Once this process is completed, the shear-induced lift by the LB's wake becomes repulsive, causing the TB to depart from the wall at a larger velocity and hence a separation of the bubble pair in the wall-normal plane.

In the following sections, we will elaborate on the key features of the two distinct escape scenarios outlined above, namely, the wall-normal escape and the wall-parallel escape. As for the head-on coalescence scenario, the evolution of the path is governed by the attractive effect of the LB's wake, similar to the corresponding unbounded condition. Given that this underlying mechanism has been comprehensively discussed in the previous work by \citet[see section 6.4 therein]{2021_zhang}, the corresponding discussion in the wall-bounded condition is omitted for the sake of brevity. To better examine the effects of pair interaction, we frequently present results from simulations conducted at the same $(\Ga, \Bo)$ but without the presence of the TB, considering only the LB. These reference results will be referred to as \q{LB-only}. While not always the case, results obtained from the pair configuration but in an unbounded flow are also presented, as they help to rationalize the wall effects. As will become clearer in the upcoming sections, the picture of pair interaction becomes more complex when vortex shedding occurs in the corresponding LB-only condition. Since this event takes place in both wall-normal and wall-parallel escapes, the discussion on these two scenarios is further divided into two subsections: one in the quasi-steady regime where no vortex shedding occurs, and the other in the unsteady regime, characterized by this vortex shedding.

\subsection{Wall-parallel escape}\label{sec:3-2wpe}
\subsubsection{The quasi-steady regime where single bubble rests close to wall}\label{sec:3-2-1wpe_r}

\begin{figure}
    \centering
    \includegraphics[height=8cm]{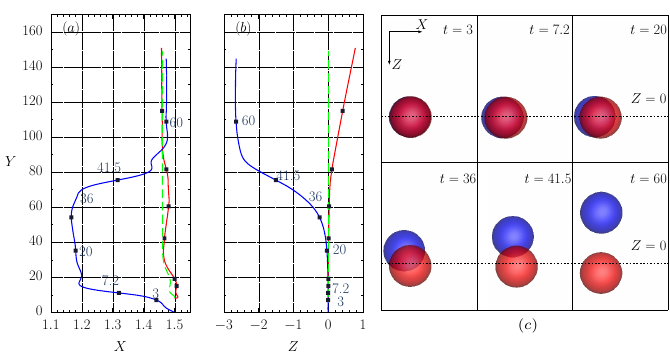}
    \caption{Path of a bubble pair following a WPE scenario ($\Ga = 15, \Bo = 0.2$). $(a)$ and $(b)$: Projections of the paths in the $(X, Y)$ and $(Z, Y)$ planes, respectively. \full{red}: LB; \full{blue}: TB; \dashed{green}: LB-only. $(c)$: Top-view positions of the two bubbles at selected time points; the wall is located at the left edge of each sub-panel; the horizontal black dashed line denotes $Z=0$.}
    \label{fig:3wpe_r_path}
\end{figure}

Figure \ref{fig:3wpe_r_path}$(a,b)$ display the rising path of a bubble pair in a typical wall-parallel escape scenario (hereafter denoted as WPE) in the $(X, Y)$ and $(Z, Y)$ planes, respectively. The case considered here is $(\Ga, \Bo) = (15, 0.2)$, corresponding to $(\Rey, \upchi) \approx (53.6, 1.17)$. In the corresponding LB-only condition, the bubble finally rests at a wall-normal position close to the wall. In the pair configuration, initially, the motion is confined within the wall-normal plane, persisting for $t$ up to about 20 (see panel $b$). During this period, the LB path is only weakly affected by the presence of the TB: it almost coincides with that in the LB-only condition. In contrast, the TB path is strongly affected by the pair interaction, continuously bending towards the wall until the TB stabilizes at a wall-normal position much closer to the wall (see panel $a$). Nevertheless, the horizontal separation between the two bubbles never exceeds 0.5, indicating that the TB remains within the wake of the LB during this stage. This 2D configuration formed in the wall-normal plane breaks down at $t\approx20$, after which the separation along $Z$ between the two bubbles starts to increase. The departure along $Z$ is highlighted in figure \ref{fig:3wpe_r_path}$(c)$, where we show the position of the two bubbles from the top view at several selected time points. As this departure begins, the TB migrates along $-Z$, while the LB is left virtually unaffected by this off-plane motion until $t\approx36$. Starting from this time, the LB is observed to continuously migrate along $Z$ (i.e., anti-parallel to the TB), while its wall-normal position slightly reverts to that observed in the LB-only condition. During the same time period, the TB is seen departing from the wall until $t\approx60$, after which it nearly stabilizes at a wall-normal position close to that of the LB.

\begin{figure}
    \centering
    \includegraphics[height=11cm]{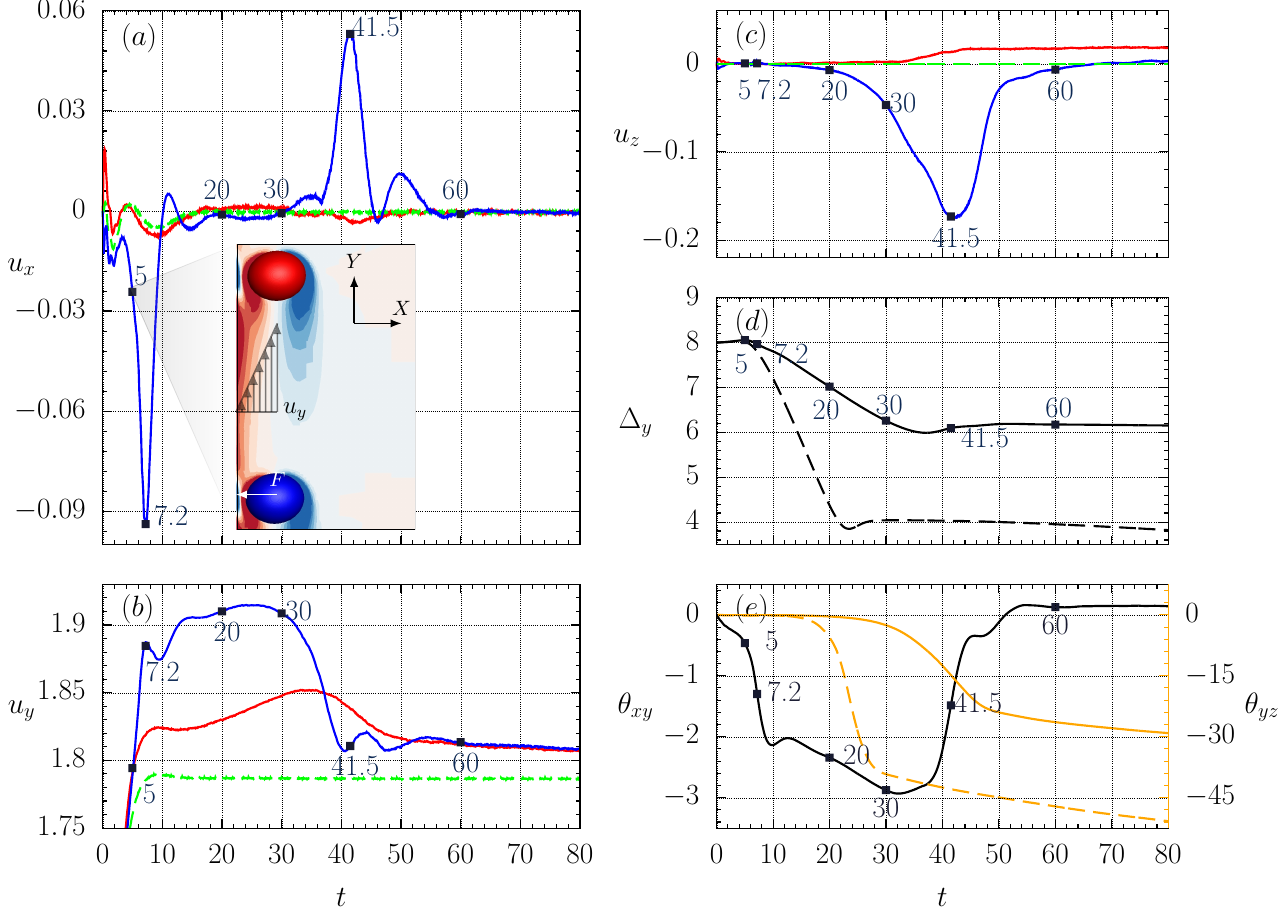}
    \caption{Evolution of selected motion characteristics for the case depicted in figure \ref{fig:3wpe_r_path}. $(a)$ Wall-normal velocity; $(b)$ Rising velocity; $(c)$ Wall-parallel velocity; Vertical separation; $(e)$ Inclination angles of $\theta_{xy}$ and $\theta_{zy}$. In panels $(a,~b,~c)$: \full{red}:~LB; \full{blue}:~ TB; \dashed{green}:~LB-only. In panel $(e)$, the black (respectively, yellow) line and tick labels represent $\theta_{xy}$ (respectively, $\theta_{zy}$). Dashed lines in panels $(d)$ and $(e)$ denote the results obtained in the corresponding unbounded condition. The inset in $(a)$ illustrates the isocontours of the spanwise vorticity $\omega_z$ at $t=5$ in the symmetry plane $z=0$. The red (resp. blue) threads represent positive (resp. negative) values, with the maximum magnitude set at 1.}
    \label{fig:4wpe_r_velo}
\end{figure}

Figure \ref{fig:4wpe_r_velo} shows the time evolution of some motion characteristics during the near-wall rising motion. In panel $(a)$, the evolution of the wall-normal velocity ($u_x$) of the LB closely resembles that in the LB-only condition, indicating that, in the pair configuration, the presence of the TB does not strongly affect the wall-normal motion of the LB. Note also that in both cases, $u_x$ of the LB remains small throughout the evolution, indicating that the net wall-normal force due to the presence of the wall remains small throughout the motion. This is not surprising. Indeed, according to the trajectories shown in figure \ref{fig:3wpe_r_path}$(a)$, the initial wall-normal position of the TB is close to that achieved in the terminal state. Hence, during the rise of the LB, the repulsive wall-normal force resulting from vortical effects \citep{1977_Vasseur, 2002_Takemura} roughly balances with the attractive one from irrotational effects \citep{1976_Wijngaarden, 1993_Kok_a}, leading the TB to rise nearly vertically upwards.
In contrast, the wall-normal velocity of the TB, $u_x^\text{TB}$, increases rapidly as $t$ increases from 3 to 7.2. The cause of this rapid increase can be inferred from the structure of the spanwise vorticity at $t=5$, shown in the inset of panel $(a)$. At this time, although the TB has not yet reached the initial vertical position of the LB, it has already entered the wall boundary layer formed between the LB and the wall (see the vorticity thread ahead of the TB in light red; for more details about the shear flow in the gap, refer to figure \ref{fig:18flowfield} in Appendix \ref{sec:append-flowfield}). Upstream of the TB, fluid elements within the wall boundary layer are entrained upwards, causing the TB to experience a shear-induced wall-normal force along $-X$. 
During the same period, the level of deformation of the two bubbles, characterized by the ratio of the major-to-minor axes $\chi$, evolve differently over time, due to pairwise interaction. Specifically, the presence of the TB modifies the flow field at the rear of the LB, increasing the front-aft pressure difference experienced by the LB. In contrast, the wake entrainment induced by the LB reduces the front-aft pressure difference of the TB. As a result, over time, the aspect ratio of the LB increases, while that of the TB decreases. These evolutions closely resemble those in the unbounded condition reported in \cite{2021_zhang} and are therefore omitted from figure \ref{fig:4wpe_r_velo}.

As the TB migrates towards the wall, the wall boundary layer close to it enhances. This amplifies the vortical interaction between the TB and the wall which leads to a repulsive contribution to the wall-normal force acting on the TB \citep{2002_Takemura}. Starting from $t \approx 7.2$, this repulsive contribution is capable in counterbalancing the attractive one induced by the shear flow ahead of the TB, and the lateral drag force makes $u_x^\text{TB}$ to quickly decrease to vanishingly small values. The vortical wall-bubble interaction also increases the dissipation rate in the gap \citep{2015_Sugioka,2020_Shi}, leading to an increase in the drag on the TB. This is why the rising velocity of the TB, $u_y^\text{TB}$, experiences a decrease within the short time interval starting at $t = 7.2$ (see figure \ref{fig:4wpe_r_velo}$b$). This drag augmentation is quickly surpassed by the attractive effect of the LB's wake, and $u_y$ of both bubbles re-increases from $t \approx 10$. Another potential factor influencing the migration of the TB is the left-right asymmetric deformation of the bubble. At $t = 10$, the maximum curvature on the wall-facing side of the TB is $\kappa = 1.25$, which exceeds the curvature on the fluid-facing side, $\kappa = 1.17$. This asymmetry arises from a more intense Bernoulli effect within the narrow gap, which has a thickness of $0.43$, leading to a pressure decrease in that region \cite{2024_Shi_b}.

For $t$ increases from 12 to 30, the vertical separation of the two bubbles gradually decreases from 7.6 to about 6.25 (see figure \ref{fig:4wpe_r_velo}$d$). Consequently, the pair interaction increases over time. Meanwhile, the TB seems to reach a quasi-equilibrium wall-normal position close to the wall: $X^\text{TB}$ slowly decreases from 1.2 to about 1.16 according to figure \ref{fig:3wpe_r_path}. To verify whether the wall-normal motion of the TB remains within the quasi-steady state, we examine the variation in the magnitude of the two distinct contributions to the wall-normal force. For the pair interaction, note that the TB has a local Reynolds number of $\Rey = 2u_y^\text{TB} \Ga \approx 57$ and $|\theta_{xy}|$ never exceeds $3^\circ$ (figure \ref{fig:4wpe_r_velo}$e$). Then, according to \citet{2011_Hallez}, the TB experiences a wall-normal force pointing along $-X$. To the leading order, this force is proportional to $\Delta_y^{-4}$ and, hence, has increased by about 2.2 times during this rising period. Conversely, the repulsive force induced by the wall-bubble interaction scales as $(X^\text{TB}-1)^{-7/2}$ \citep{2024_Shi_a} when the wall separation, $X^\text{TB}-1$, is of $O(0.1)$. This indicates that the repulsive wall-normal force contributed from the wall-bubble interaction has also increased by about 2.2 times during the same period. Integrating these observations, we conclude that the wall-normal motion of the TB remains in a quasi-steady state, and transient effects do not play a key role at this stage. 

Let us now examine the motion in the second stage, where the symmetry configuration breaks down and the TB escapes from the LB's wake in the wall-parallel plane. For an unbounded bubble pair with small-to-moderate shape deformation \citep{2021_zhang}, the in-line configuration is known to break down when the difference $u_y^\text{TB} - u_y^\text{LB}$ formed during the pair interaction process exceeds a critical, case-dependent value. This feature seems to persist in the wall-bounded condition, as the separation along $Z$ starts to grow at $t \approx 20$, up to which $u_y^\text{TB} - u_y^\text{LB}$ increases to about 0.085. However, unlike in the unbounded condition where the escape can occur in any vertical plane depending on the growth of disturbance, the repulsive transverse force from the vortical wall-bubble interaction makes the TB's escape in the wall-normal plane difficult. In contrast, escape in the wall-parallel plane is more feasible. 
Notably, once the TB, moving within the wake of the LB, has an initial offset from the symmetry plane $Z=0$, it encounters a shear flow. Within the LB's wake, the velocity gradient $\partial_z U_y$ is $<0$ (respectively, $>0$) for $z>0$ (respectively, $<0$). Consequently, the shear-induced sideways force \citep{1987_Auton,1998_Legendre} tends to deviate the TB from the symmetry plane in the direction that amplifies the initial offset. This is why at $t\approx20$, the spanwise velocity of the TB, $u_z^\text{TB}$, starts to increase (see figure \ref{fig:4wpe_r_velo}$c$). This escape also distorts the wake of the LB. Consequently, at a later time of about $t\approx30$, the LB begins to depart from the symmetry plane with a spanwise velocity that is anti-parallel to $u_z^\text{TB}$.

As the TB starts to escape, the misalignment of the bubble pair in the $(Z, Y)$ plane grows rapidly (see $\theta_{zy}$ in figure \ref{fig:4wpe_r_velo}$e$). Specifically, $\theta_{zy}$ becomes an order of magnitude larger than $\theta_{xy}$ at $t \approx 41.5$, indicating that the pair interaction is largely within the $(Z, Y)$ plane in the subsequent process. Consequently, the vortical wall-bubble interaction drives the TB away from the wall (see $u_x^\text{TB}$ in figure \ref{fig:4wpe_r_velo}$a$) until $t \approx 60$ when the TB reaches a new equilibrium position (see figure \ref{fig:3wpe_r_path}$a$). Another consequence of the escape is that the attractive effect of the LB's wake is strongly attenuated \citep{2011_Hallez}. This is why, according to figure \ref{fig:4wpe_r_velo}$(c)$, the rising velocity of the two bubbles experiences a decrease during the escape. Note, however, that even beyond $t \approx 60$, the separation in the $(Z, Y)$ plane is not entirely saturated. Specifically, $u_z^\text{LB}$ remains at small but positive values, causing $\theta_{zy}$ to continue to grow over time. 

The aspect ratio of both bubbles returns to the value under LB-only condition at final stage after $t=60$. However the final geometry of the bubble pair differs significantly from that in the corresponding unbounded condition. To highlight this difference, results for the vertical separation and the inclination angle obtained in the unbounded condition at the same $(\Ga, \Bo)$ are also shown in figure \ref{fig:4wpe_r_velo} for comparison. The terminal vertical separation is about 6.2 in the wall-bounded configuration, while it is only 4 in the unbounded condition (figure \ref{fig:4wpe_r_velo}$d$). Furthermore, at $t=80$—the longest duration over which the simulations are carried out—the inclination angle nearly saturates at $29^\circ$ in the presence of wall (figure \ref{fig:4wpe_r_velo}$e$), whereas it is about $50^\circ$ in the unbounded condition (close to the critical value $\theta_{zy}^c \approx 53^\circ$ at which the repulsive pair interaction vanishes \citep{2011_Hallez}). These drastically different behaviors are linked to the wake effect of the LB, which is strongly attenuated by the presence of the wall \citep{2003_Magnaudet,2020_Shi}. This attenuation can be clearly seen by examining the evolution in $\Delta_y(t)$ prior to its saturation. Given that the difference $u_y^\text{TB} - u_y^\text{LB}$ originates largely from the attraction of the LB's wake, the attenuation corresponds to a smaller slope in $\Delta_y(t)$ in the presence of the wall, which is confirmed by figure \ref{fig:4wpe_r_velo}$(b)$. This attenuation also affects the evolution of the inclination angle. Specifically, during the escape, the shear flow experienced by the TB is weaker in the wall-bounded condition, owing to the larger vertical separation. This causes the corresponding inclination angle to increase more slowly and to saturate at a relatively lower value, as shown in figure \ref{fig:4wpe_r_velo}$(e)$.

\subsubsection{The unsteady regime where single bubble bounces repeatably close to wall}\label{sec:3-2-2wpe_b} 

\begin{figure}
    \centering
    \includegraphics[height=8cm]{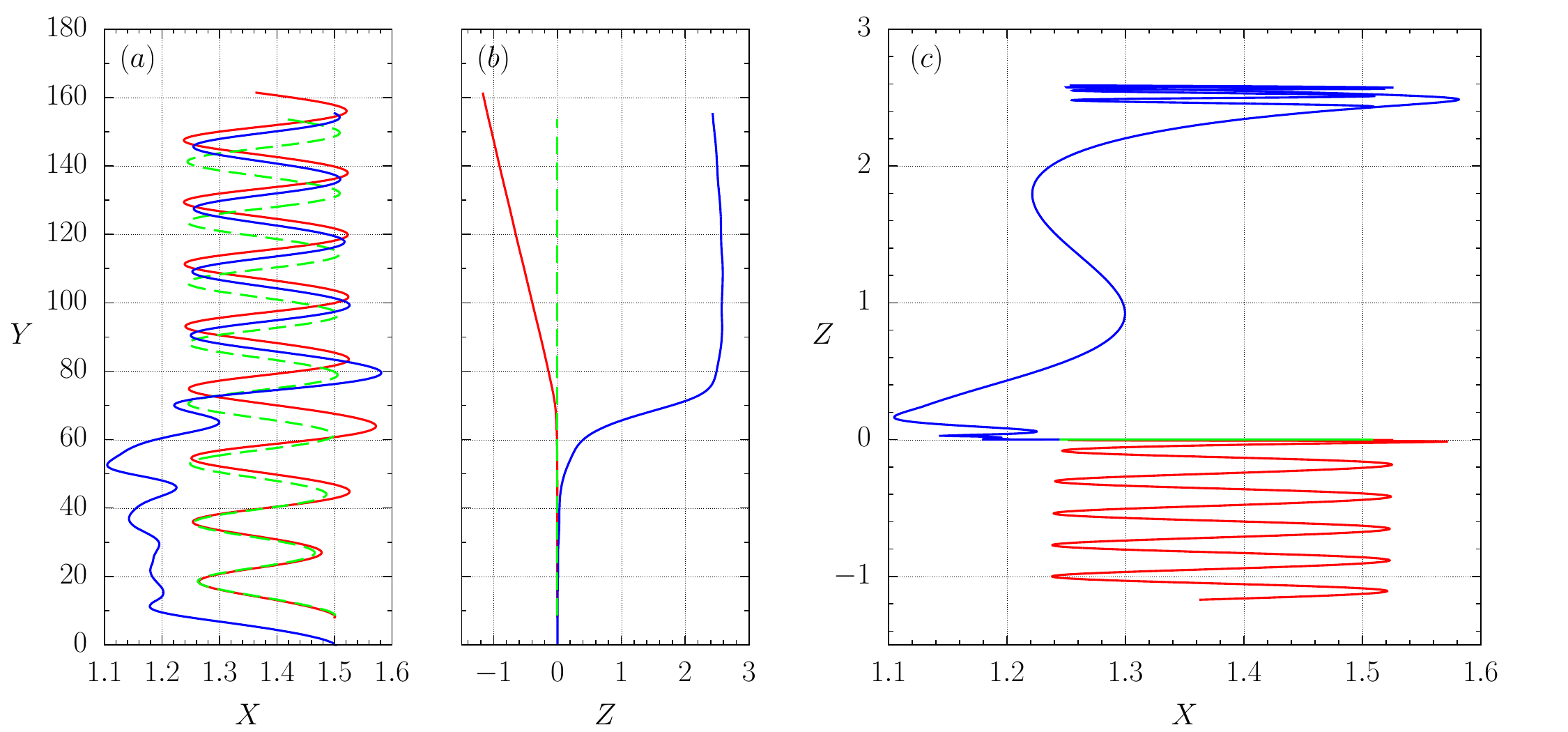}
     \caption{Path of a bubble pair following a WPE scenario in the unsteady flow regime ($\Ga = 20, \Bo = 0.2$). $(a)$, $(b)$ and $(c)$: Projections of the paths in the $(X, Y)$, $(Z, Y)$ and $(Z, X)$ planes, respectively. \full{red}: LB; \full{blue}: TB; \dashed{green}: LB-only. }
    \label{fig:5wpe_b_path}
\end{figure}

\begin{figure}
    \centering
    \includegraphics[height=11cm]{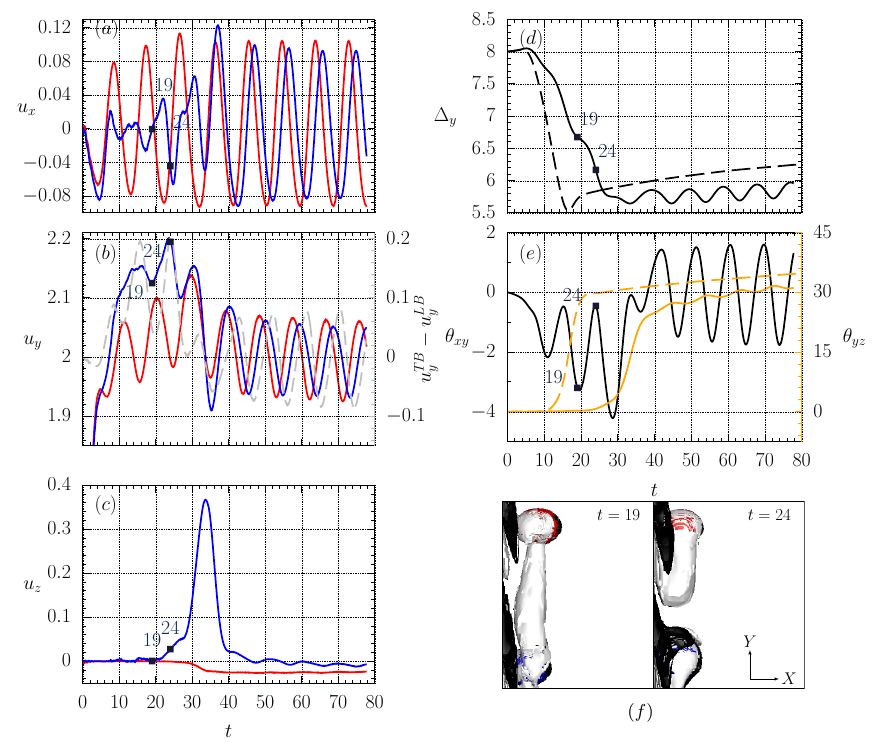}
    \caption{Evolution of selected motion characteristics for the case depicted in figure \ref{fig:5wpe_b_path}. For the correspondence of panels $(a-e)$, see caption in figure \ref{fig:4wpe_r_velo}. In panel $(b)$, the gray dashed line denotes the velocity difference $u_y^{TB} - u_y^{LB}$, with the tick marks corresponding to the right $y$-axis. Panel $(f)$ shows the isosurface $\omega_y=\pm 0.1$ of the streamwise vorticity at selected time points for the case $(\Ga,\Bo)=(20,0.2)$. The black (respectively, light gray) thread corresponds to negative (respectively, positive) values. }
    \label{fig:6wpe_b_velo}
\end{figure}

The WPE scenario discussed in the previous section can also manifest in the unsteady regime featured by vortex shedding, provided that $\Bo$ remains below a critical value that increases with increasing $\Ga$ (see figure \ref{fig:2phase}$a$). An example of the WPE scenario in the unsteady regime is depicted in figure \ref{fig:5wpe_b_path}$(a, b)$, which show the path of the bubble pair in the $(X,Y)$ and $(Z,Y)$ planes, respectively. The case considered is $(\Ga, \Bo) = (20, 0.2)$, corresponding to $(\Rey,\upchi) \approx (79, 1.22)$. As a supplement, figure \ref{fig:6wpe_b_velo} illustrates the time evolution of selected motion characteristics. According to figure \ref{fig:5wpe_b_path}$(a)$, the bubble exhibits a repeatable near-wall bouncing motion in the LB-only condition. In the pair configuration, the motion of the bubble pair remains confined to the wall-normal plane until $t \approx 24$, at which $u_y^\text{TB}-u_y^\text{LB}$ reaches a local maximum of about 0.2, as depicted in figure \ref{fig:6wpe_b_velo}$b$.

Prior to the escape, the wall-normal motion of the TB is primarily confined to the near-wall region where $X \lesssim 1.2$. Conversely, the LB exhibits sizable wall-normal oscillations in the region where $X > 1.25$, almost in identical with that of the LB-only condition. The suppression of the lateral oscillation of the TB, compared with that in the LB-only condition, may be understood by noting that the elongated boundary layer of the LB attenuates the negative vortical thread in the gap separating the TB and the wall (see, e.g., the inset in figure \ref{fig:4wpe_r_velo}$a$). This vortical thread is closely linked with the lateral oscillation. Specifically, in the LB-only condition, repeatably near-wall oscillations appear only when the maximum of these vortical threads exceeds a critical value of around 6 \citep{2024_Shi_b}. 
The lateral oscillations of the LB are accompanied by strong vortex shedding, as depicted in figure \ref{fig:6wpe_b_velo}$(f)$, which shows the structure of the streamwise vorticity in the wake of the LB. Given that these vortex cores correspond to low-pressure regions, when these shed vortices are advected downstream, the drag on the TB is decreased as the vortices reach the front of the TB and then increased as they reach the TB's wake. This is primarily the reason that the difference $u_y^\text{TB} - u_y^\text{LB}$ oscillates during this period. Given the repulsive nature of the pair interaction, the maximum wall-normal separation $X^\text{LB}-X^\text{TB}$ achieved in each round of LB's oscillation grows with each oscillation cycle. Owing to this growth, one might expect the separation of the bubble pair to take place in the wall-normal plane after a sufficient number of oscillation rounds. Unfortunately, this does not occur as the TB begins to depart along $Z$ just after the third round of oscillation. 

The wall-parallel separation of the bubble pair accompanied with strong wall-normal oscillations is highlighted in figure \ref{fig:5wpe_b_path}$(c)$, which shows the path of the bubble pair in the $(Z, X)$ plane. The separation nearly saturates at $t \approx 40$ (although $u_z^\text{LB}$, according to figure \ref{fig:6wpe_b_velo}$(d)$, remains at small but negative values), beyond which the two bubbles exhibit wall-normal oscillations with fixed frequency and amplitude. Interestingly, while the frequencies of the oscillations are almost identical, there is a phase difference of about $T/4$ (with $T$ denoting the period of the wall-normal oscillation), as seen from the evolutions of $u_x$ and $u_y$. This phase difference causes $\theta_{xy}$ to oscillate over time, instead of remaining at a vanishingly small value as in a quasi-steady WPE scenario (see, e.g., $\theta_{xy}$ in figure \ref{fig:4wpe_r_velo}).

To wrap up this section, we examine the effects of the wall on the evolution of the pair configuration. In figure \ref{fig:6wpe_b_velo}, we include results for the vertical separation and the inclination angle obtained from the unbounded condition at the same $(\Ga, \Bo)$ for comparison. In the absence of the wall, the escape occurs earlier, consistent with the WPE scenario in the quasi-steady regime. As the escape reaches saturation, the final vertical separation and inclination angle are approximately $(6, 31.5^\circ)$ in the presence of the wall and about $(6.25, 35^\circ)$ in the unbounded condition, indicating only marginal effects of the wall. This contrasts significantly with the quasi-steady WPE scenario, where the wall notably reduces the final vertical separation and increases the final inclination. These differences are not surprising, as the suppression of wake development by the wall becomes insignificant once the bubble pair begins to bounce. Consequently, even with the wall's presence, the final pair configuration in an unsteady WPE scenario closely resembles that found in the unbounded condition.

\subsection{Wall-normal escape}\label{sec:3-3wne}

\subsubsection{The unsteady regime where single bubble bounces repeatably close to wall}\label{sec:3-3-1wne_b}

The unsteady WPE scenario discussed in \S~\ref{sec:3-2-2wpe_b} does not persist if the TB gains a finite departing velocity capable of overcoming the wall-ward attraction by the LB's wake. Once this process is completed, the TB reaches the unbounded side of the LB's wake, where the shear-induced lift enables the TB to escape from the LB's wake in the wall-normal plane. Hereafter, this type of escape will be termed the (TB-led) unsteady WNE scenario. According to figure~\ref{fig:2phase}($a$) (solid cycle symbols), this scenario occurs in the regime characterized by large $\Ga$ and moderate $\Bo$.

\begin{figure}
    \centering
    \includegraphics[height=8cm]{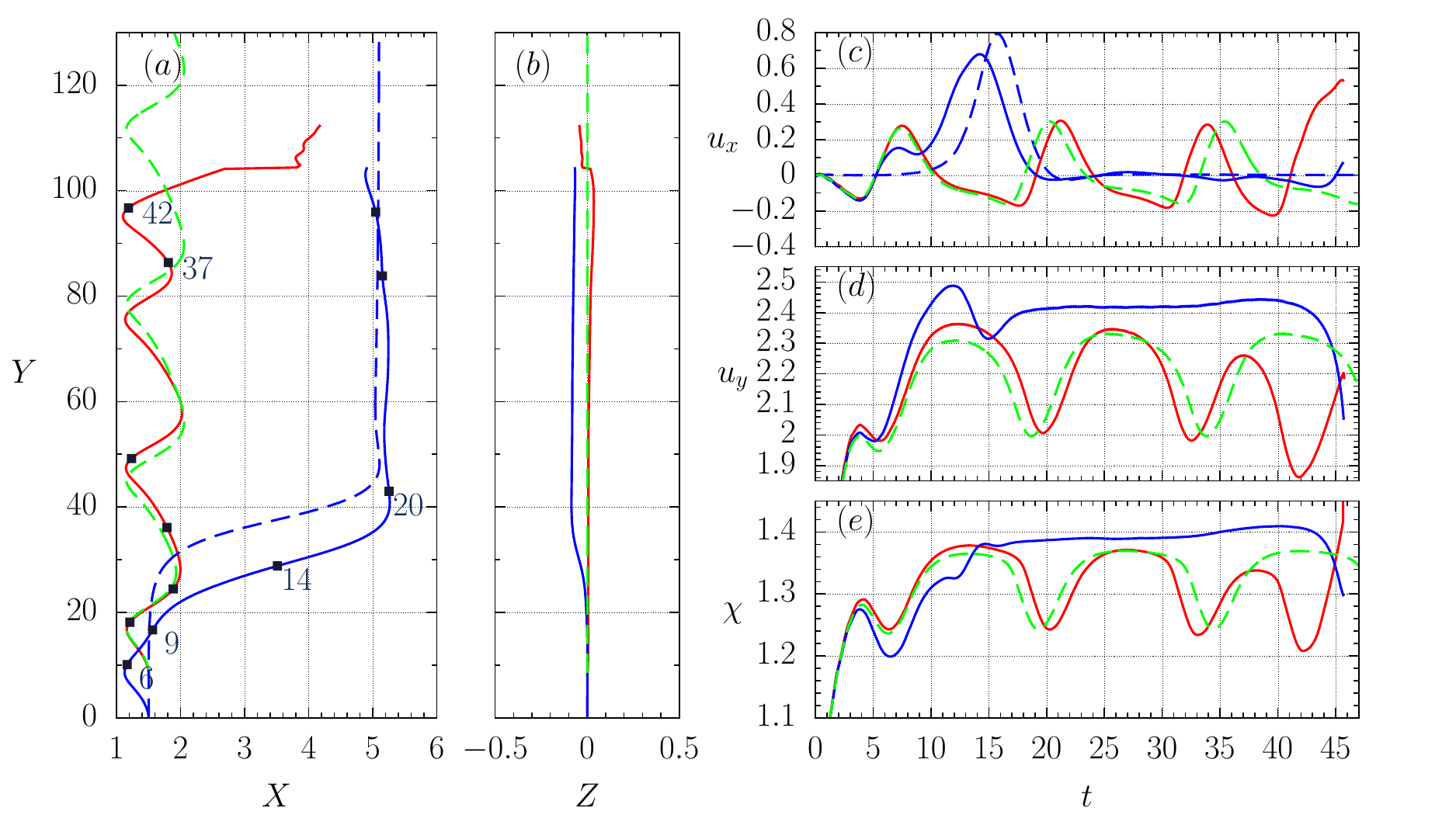}
    \caption{Path and selected motion characteristics of a bubble pair following an unsteady, TB-led WNE scenario ($\Ga = 25, \Bo = 0.2$). $(a)$ and $(b)$: Projections of the paths in the $(X, Y)$ and $(Z, Y)$ planes, respectively. $(c)$: Wall-normal velocity; $(d)$: Vertical velocity; $(e)$: Aspect ratio. In all panels, \full{red}: LB; \full{blue}: TB; \dashed{green}: LB-only. In $(a)$ [respectively, $(c)$], \dashed{blue} represents the path (respectively, wall-normal velocity) of the TB in the unbounded condition at the same $(\Ga, \Bo)$. }
    \label{fig:7wne_b_path}
\end{figure}

Figure \ref{fig:7wne_b_path}$(a,~b)$ shows the path of a bubble pair following an unsteady WNE scenario. The selected case is $(\Ga, \Bo) = (25, 0.2)$, corresponding to $(\Rey,\upchi)=(111,1.31)$. Throughout the motion, the separation along $Z$ of the two bubbles never exceeds $0.1$, indicating that they do not separate from each other in the wall-parallel plane. Conversely, their path in the $(X, Y)$ plane reveals that, after the first near-wall bouncing at $t \approx 6$, the TB continuously migrates away from the wall, reaching a wall-normal position of $X^\text{TB}\approx5.2$ at $t = 20$. Given that $X^\text{LB}$ never exceeds $2.0$ during this period, the TB effectively escapes from the LB's wake. This behavior of the TB contrasts markedly with that observed in the initial stage of an unsteady WPE scenario, where the TB remains \q{trapped} in the gap between the LB and the wall prior to the escape along $Z$ (see figure \ref{fig:5wpe_b_path}$a$).

\begin{figure}
    \centering
    \includegraphics[height=4.0cm]{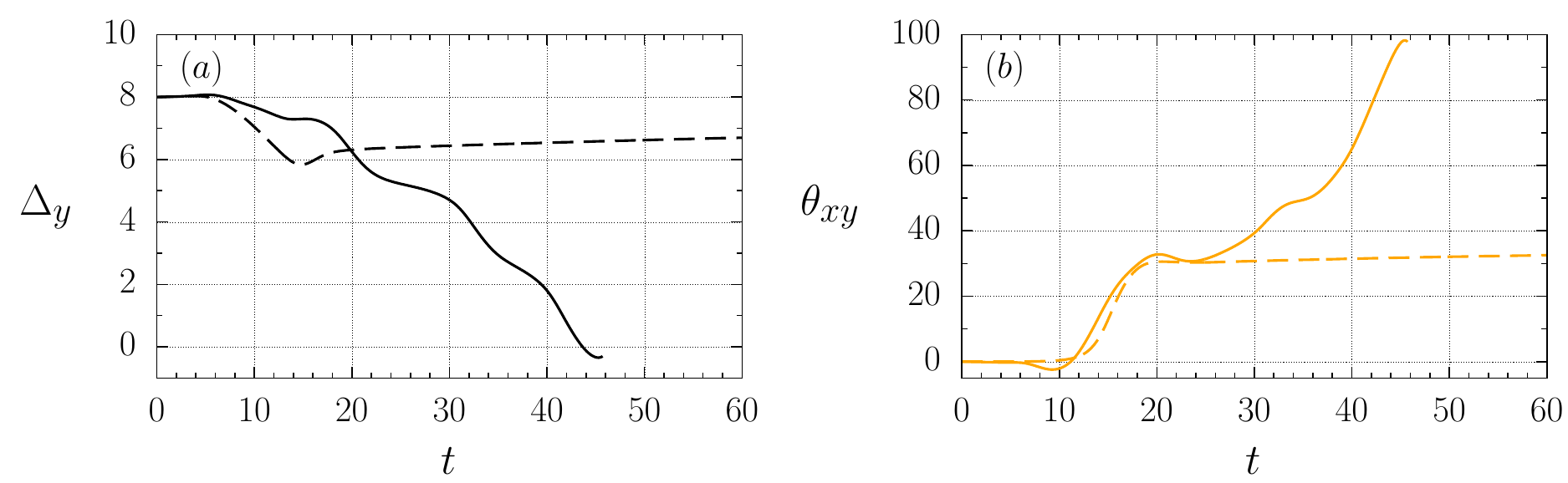}
    \caption{Evolution of $(a)$ the vertical separation and $(b)$ the inclination angle of the bubble pair depicted in figure \ref{fig:7wne_b_path}. In both panels, the solid and dashed lines represent results in the wall-bounded and the unbounded conditions, respectively. Note that the plots in the wall-bounded condition terminate at $t \approx 45$, as the two bubbles coalesce at this time.}
    \label{fig:8wne_b_geo}
\end{figure}

Figure \ref{fig:7wne_b_path}$(c,~d)$ and figure \ref{fig:8wne_b_geo} present the evolution of the velocities and relative position of the bubble pair, respectively. Notably, $u_x(t)$ for the two bubbles does not exhibit significant differences for $t$ up to about 6. This suggests that the pair interaction, which, in the initial stage of a WPE scenario, tends to drive the TB faster towards the wall (see figure \ref{fig:4wpe_r_velo}$a$), does not substantially influence the motion of the TB here. The \emph{silence} of this effect can be attributed to the fact that the maximum $u_x$ of the LB during the initial approaching stage is $-0.18$, twice that shown in figure \ref{fig:4wpe_r_velo}$(a)$. Hence, for this bubble pair, the attractive contribution to the wall-normal force is largely governed by the irrotational effect. Nevertheless, the signature of the attractive effect by pair interaction can still be discerned by noting that the closest wall distance reached by the TB is $X=1.13$, smaller than the value $1.16$ for the LB. Considering that the wall-normal force from the wall-bubble interaction is proportional to $(X-1)^{-7/2}$, at the closest wall position, the TB experiences a repulsive wall-normal force twice that acting on the LB. This migration closer to the wall is clearly a consequence of the pair interaction. 

According to figure \ref{fig:7wne_b_path}$(a)$, the TB reaches the initial vertical position of the LB at $t\approx6$. At this time point, the TB is located on the wall-bounded side of the LB's wake and begins to experience a strong shear-induced lift pointing towards the wall, causing the difference $u_x^\text{LB}-u_x^\text{TB}$ to grow. This process does not last long, as the finite positive $u_x^\text{TB}$ makes the TB  to quickly reach the unbounded side of the LB's wake. The wall-normal escape begins at $t \approx 9$, when $u_x^\text{TB}$ exceeds that of the LB. Subsequently, $u_x^\text{TB}$ rapidly increases, peaking at about 0.7 at $t=14$. Afterwards, it begins to decrease and stabilizes at approximately zero at $t \approx 20$. Throughout the period of $9\leq t \leq20$, the TB remains on the unbounded side of the LB, indicating that the shear-induced lift promotes the TB's wall-normal escape. However, it is important to note that the escape occurs earlier than in the corresponding unbounded condition, due to the small but finite wall-normal velocity of the TB formed at $t\approx6$. In other words, the wall now promotes the escape, instead of delaying its occurrence as in a WPE scenario.

To highlight this promoted escape, we show in figure \ref{fig:7wne_b_path}$(a)$ the path of the TB in the escape plane in the corresponding unbounded condition (the initial position of this path is shifted so that it starts at the same position as in the wall-bounded condition). Comparing the two TB paths, it seems that an earlier escape results in a larger horizontal drift of the TB. This trend is surprising at first, as an earlier escape corresponds to a weaker shear flow experienced by the TB, confirmed by the larger vertical separation in the wall-bounded condition shown in figure \ref{fig:8wne_b_geo}$(a)$. This is known \citep{2021_zhang} to lead to a decrease in the final lateral drift of the TB. Moreover, during the escape, the TB also experiences an irrotational, attractive force from the wall when the wall distance surpasses the mean position in the LB-only condition. We find the larger horizontal drift of the TB identified here to be closely linked to the interaction between the TB and the streamwise vortices shed from the LB, a distinct feature in an unsteady, TB-led WNE scenario that will be detailed in the following.

\begin{figure}
    \centering
    \includegraphics[height=8cm]{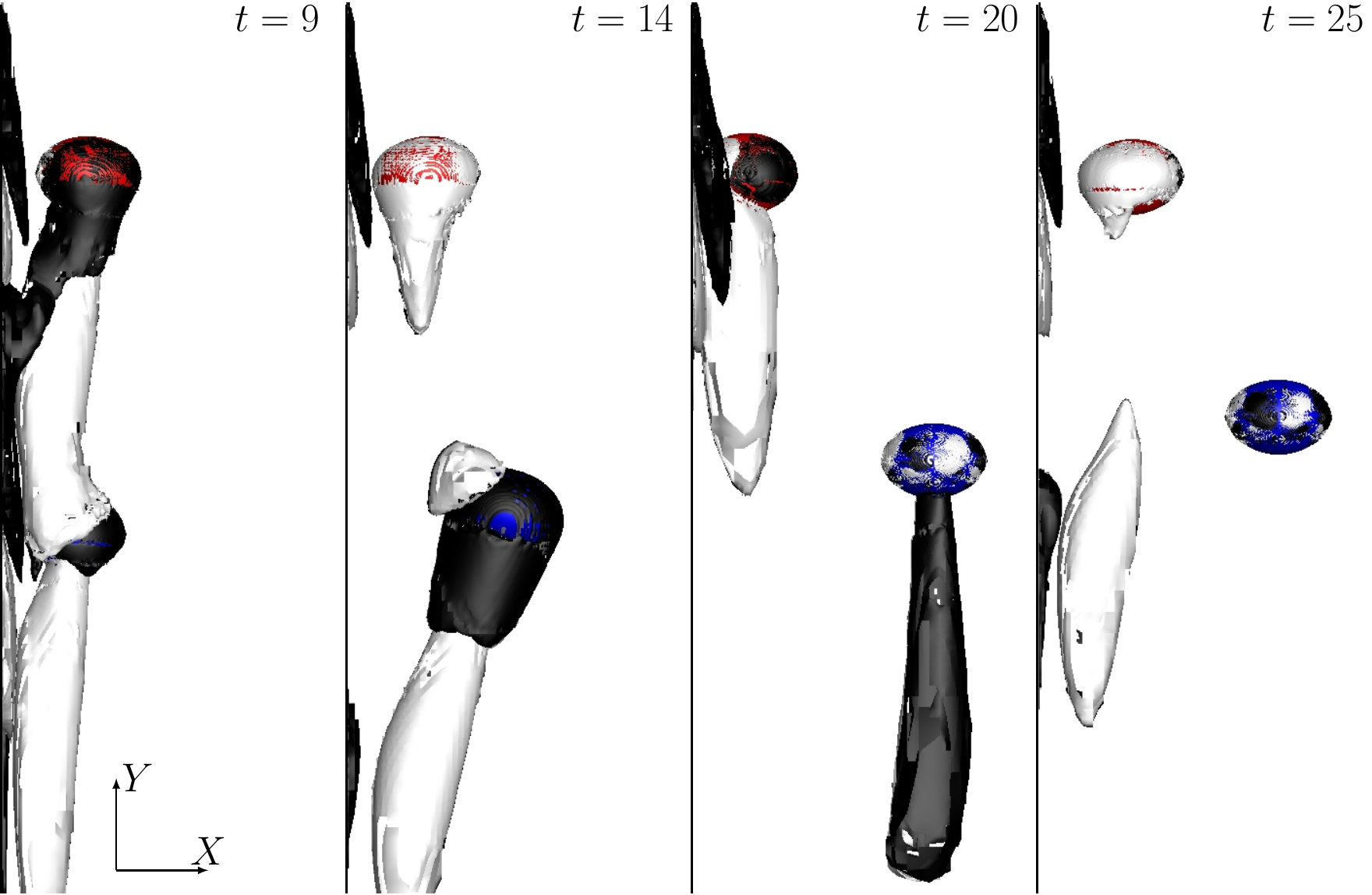}
    \caption{Isosurface $\omega_y=\pm 0.1$ of the streamwise vorticity at selected time points for the case $(\Ga,\Bo)=(25,0.2)$. The black (respectively, light gray) thread corresponds to negative (respectively, positive) values. }
    \label{fig:9wne_b_vor}
\end{figure}

To begin, we illustrate in figure \ref{fig:9wne_b_vor} the structure of the streamwise vorticity $\omega_y$ at several selected time frames during the escape. At $t = 9$, two vortical threads are visible in the wake of each bubble (note, however, that only one of the two threads is shown, as the flow remains symmetric with respect to $Z=0$). Hereafter, the pairs of threads associated with the LB and the TB will be referred to as $\omega_y^\text{LB}$ and $\omega_y^\text{TB}$, respectively. 
As $t$ increases from 9 to 14, $\omega_y^\text{LB}$ are advected further downstream, becoming rather weak upstream of the TB. Meanwhile, the intensity of the major vortical threads of $\omega_y^\text{TB}$ (colored in light gray) gradually increases. This intensification can be explained by examining the budget equation for $\omega_y^\text{TB}$. Specifically, as $\omega_y^\text{LB}$ reach the wake of the TB, they induce a stretching/tilting term, $\omega_y^\text{LB} \cdot \partial U_y / \partial y$, in the $\omega_y^\text{TB}$ budget \citep{2009_Adoua}, where $U_y$ is the liquid velocity along $Y$ and $\partial U_y / \partial y$ is the extensional rate of the liquid flow. Because $\partial U_y / \partial y$ is positive irrespective of $Z$ in the wake of the TB, $\omega_y^\text{TB}$ is amplified as their orientations align with those of $\omega_y^\text{LB}$. Since the pair of streamwise vortices in the wake are directly responsible for the repulsive wall-normal force acting on the bubble \citep{1987_Auton,2024_Shi_a}, an increase in their intensity leads to a rapid increase in $u_x^\text{TB}$. This is primarily the reason that the lateral drift of the TB is larger than in the unbounded condition where no such interplay between the TB and the vortices shed from the LB takes place. 

In addition to the amplification by $\omega_y^\text{LB}$, the generation of $\omega_y^\text{TB}$ is also influenced by the azimuthal surface vorticity $\omega_\phi^\text{TB}$. As revealed from figure \ref{fig:7wne_b_path}$(c, d)$, at $t \approx 14$, the TB migrates away from the wall at an angle of $\tan^{-1}(u_x/u_y) \approx 15^\circ$ with respect to the vertical. Consequently, the projection of $\omega_\phi^\text{TB}$ along $Y$ is non-negligible at this moment. This results in a pair of vortical threads with orientations that are anti-parallel to those of the primary ones amplified by $\omega_y^\text{LB}$. In figure \ref{fig:9wne_b_vor}, this pair of vortices appears as a black vortical thread surrounding the TB. As $t$ increases from 14, the amplification by $\omega_y^\text{LB}$ decays. Meanwhile, $u_x^\text{TB}$ decreases over time. Consequently, the intensities of the two distinct vortical pairs decrease over time: the primary one becomes indistinguishable by $t \approx 20$, while the one arising from the azimuthal surface vorticity vanishes by $t \approx 25$.

Coming back to the motion of the bubble pair, during the period when $22 < t < 32$, the interaction between the two bubbles remains subtle. In particular, $u_y^\text{TB}$ shows only a weak dependence on $t$ and closely aligns with the terminal velocity observed in the unbounded, isolated condition \citep{2021_zhang}. During the same period, the LB undergoes a regular, near-wall bouncing motion as in the LB-only condition. However, this situation does not persist, as the wall-induced drag increase causes the LB to rise more slowly than the TB, thereby decreasing their vertical separation over time (see $\Delta_y$ in figure \ref{fig:8wne_b_geo}$a$). By $t = 32$, $\Delta_y$ is less than $4$, indicating that the pair interaction becomes significant in the subsequent processes.

At $t\approx33$, the inclination angle $\theta_{xy}$ according to figure \ref{fig:8wne_b_geo}$(b)$ increases to about $50^\circ$. This is roughly the critical angle at which the pair interaction switches from repulsive to attractive \citep{2011_Hallez}. Hence, in the subsequent processes, the TB is attracted towards the LB, causing $u_x^\text{TB}$ to remain negative (although small) during $33\lesssim t \lesssim45$. The influence on the LB's motion is more complex, as the wall effects also play a significant role. Specifically, the pair interaction enhances the drag on the LB, thereby decreasing $u_y^\text{LB}$ and the local slip Reynolds number. Given this decrease, the history effects become less significant \citep{2024_Shi_b}, and the oscillation of the LB is suppressed such that the LB bounces at a smaller lateral amplitude and a mean position closer to the wall (see figure \ref{fig:7wne_b_path}$a$). This suppression persists up to $t\approx44$, when the TB catches up to the LB. From this moment, the attractive force between the pair interaction, owing to the irrotational effect \cite{2019_zhang},  helps the LB to move away from the wall and eventually coalesce with the TB at $t \approx 47$. This final behavior of the bubble pair differs from that in the unbounded condition, where the TB never catches up, and the two bubbles eventually reach a stable configuration with a vertical separation and an angle of inclination of about $(7, 33^\circ)$, as seen from figure \ref{fig:8wne_b_geo}.

In addition to the path evolution of the bubble pair, another interesting aspect worth discussing is the bubble deformation, the time evolution of which is shown in figure \ref{fig:7wne_b_path}($e$). The level of deformation is largely influenced by the dynamic pressure in the vicinity of the bubble, and hence the bubble's rising speed $u_y$. Due to this dependence, the evolution of the aspect ratio closely resembles that of $u_y$, as revealed by comparing panels ($d$) and ($e$) in figure \ref{fig:7wne_b_path}. Of particular note is the evolution of the aspect ratio of the TB ($\chi^\text{TB}$) during the early escaping stage, i.e., for approximately $t$ increasing from 5 to 14. During this period, the rising speed of the TB is greater than that of the LB, while its aspect ratio is smaller. This 'abnormal' behaviour is related to the entrainment effect of the LB's wake, which allows the TB to deform less than the LB at the same rising speed.
\begin{figure}
    \centering
    \includegraphics[height=8cm]{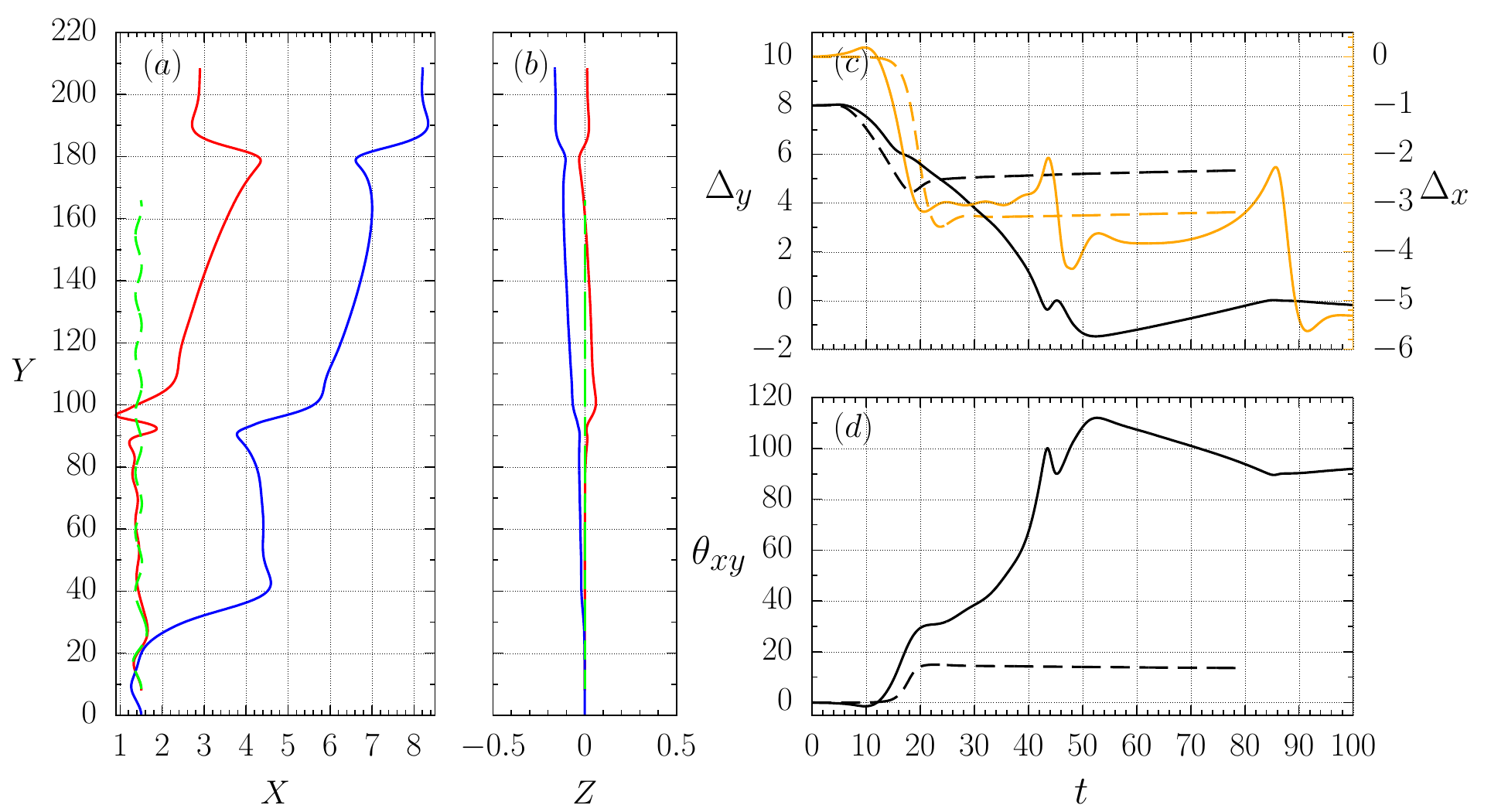}
    \caption{Path and selected motion characteristics of a bubble pair following an unsteady, TB-led WNE scenario ($\Ga = 25, \Bo = 0.3$). In panels ($a,b$), \full{red}: LB; \full{blue}: TB; \dashed{green}: LB-only. In panels $(c,d)$, the solid and dashed lines represent results in the wall-bounded and the unbounded conditions, respectively.}
    \label{fig:10wne_b_nocoal}
\end{figure}

Note, however, that not all bubble pairs following an unsteady WNE scenario will eventually coalesce. Specifically, while the wall promotes the TB to escape and eventually catch up with the LB, whether the two bubbles will ultimately coalesce depends on their state of motion when they collide \citep{1998_Duineveld, 2009_Sanada, 2017_Tripathi, 2019_Kong}. Once the amount of surface vorticity accumulated on the bubble surface exceeds a critical, case-dependent value, the two bubbles will bounce off instead of coalescing \citep{2019_zhang}. Given the close relation between surface vorticity and the extent of bubble deformation (hence $\Bo$) \citep{2007_magnaudet}, we checked the path of the bubble pair with $(\Ga, \Bo) = (25, 0.3)$, namely the case with identical $\Ga$ to that illustrated in figure \ref{fig:7wne_b_path} but with a slightly larger $\Bo$ (hence $\upchi$). Figure \ref{fig:10wne_b_nocoal} illustrates some motion characteristics of this case. After escaping (see panel $a$), the TB catches up with the LB at $t\approx42$. Afterwards, it \q{kisses} the LB twice, first at $t \approx 45$ and again at $t \approx 88$, but fails to coalesce with the LB. After the second kiss, the two bubbles bounce off and rise almost side by side (see $\theta_{xy}$ in figure \ref{fig:10wne_b_nocoal}$d$). These observations align with those reported in \citet{2019_zhang}, where the coalescence of a bubble pair with $\Rey=O(100)$ no longer occurs when $\Bo$ increases from 0.2 to 0.3.

Owing to the finite size of the computational domain, the simulation for the above case was terminated at $t=100$, a time at which point the pair was already close to the top boundary. Nevertheless, it can be anticipated that in the subsequent processes, the LB will escape from the near-wall region. This is due to the fact that at $t=100$, the wall-normal separation of the LB is $X^\text{LB} \approx 3$, which is larger than that from the \q{symmetry} plane of the pair interaction (about $\Delta_x/2 = 2.65$). Hence, the LB experiences a larger attractive force from the TB than from the wall. Given the sizable deformation of the bubble pair, we anticipate the bubble pair to finally reach an equilibrium horizontal separation and to rise side-by-side, like in a DKT scenario \citep{2021_zhang}. The path evolution here differs from that in the corresponding unbounded condition. As seen from the dashed lines in figure \ref{fig:10wne_b_nocoal}($c,d$), there, while the two bubbles still separate laterally, the TB never catches up, as part of its potential energy is spent to escape from the LB's wake. This difference, together with the similar one indicated in the case with $(\Ga, \Bo) = (25, 0.2)$, reveals a key role of the wall in an unsteady WNE scenario: it decreases the rising speed of the LB such that the TB is able to catch up and, if the bubble pair does not coalesce, promotes the bubble pair to migrate away from the wall while rising side-by-side.

\subsubsection{The quasi-steady regime where single bubble migrates away from the wall}\label{sec:3-3-2wne_d}

The repeatable near-wall bouncing motion no longer persists when the vortical effect dominates the wall-bubble interaction. In the pair configuration, while both bubbles now migrate away from the wall, the wall-induced asymmetry causes the departing velocities of the bubble pair to differ, leading to a wall-normal escape of the TB from the LB's wake. This escape scenario, occurring in the quasi-steady flow regime, will be termed the quasi-steady WNE scenario.

\begin{figure}
    \centering
    \includegraphics[height=8cm]{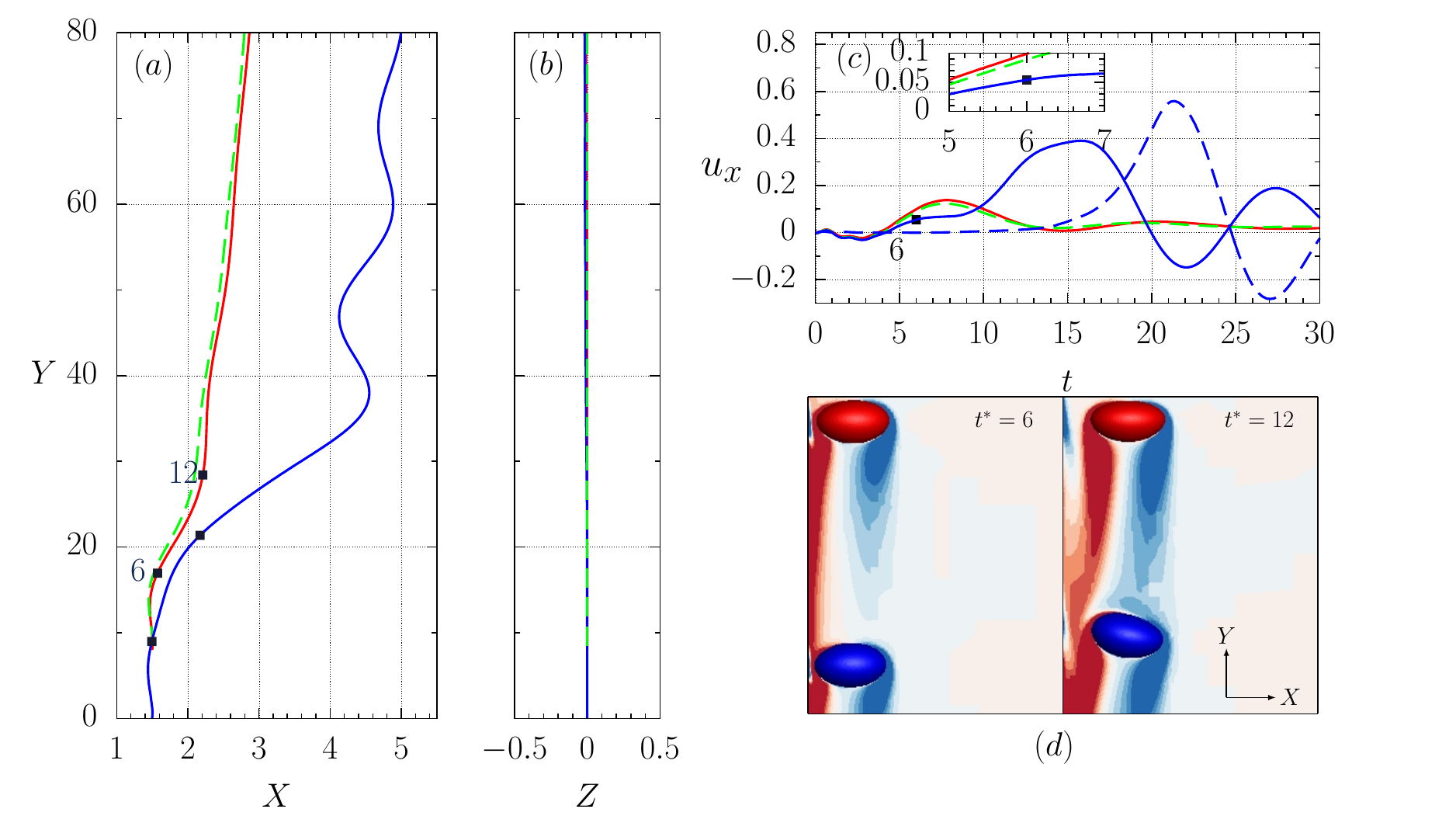}
    \caption{Path and selected motion characteristics of a bubble pair in a quasi-steady, TB-led WNE scenario ($\Ga = 30, \Bo = 0.5$). $(a)$ and $(b)$: Projections of the paths in the $(X, Y)$ and $(Z, Y)$ planes, respectively. $(c)$: Wall-normal velocity. The departing velocity is highlighted in $(c)$ which reads $u_d^{TB}=0.052$. $(d)$: Isocontours of the spanwise vorticity $\omega_z$ in the symmetry plane $z=0$. In $(d)$, the red (resp. blue) threads represent positive (resp. negative) values, with the maximum magnitude set at 1. For the correspondence of differently colored lines in $(a-c)$, refer to figure \ref{fig:7wne_b_path}.}
    \label{fig:11wne_d_TB}
\end{figure}

An example of a bubble pair undergoing a quasi-steady WNE scenario is illustrated in figure \ref{fig:11wne_d_TB}$(a, b)$, which displays the path of a bubble pair for $t$ up to about 30. The case considered is $(\Ga, \Bo) = (30, 0.5)$, corresponding to $(\Rey, \upchi) = (118, 1.76)$ in the LB-only condition. As the pair interaction starts to be effective (at $t \approx 6$), the TB stays on the wall-bounded side of the LB's wake, experiencing a shear-induced lift towards the wall. Nevertheless, the finite departing velocity of the TB at this moment (see $u_x$ in figure \ref{fig:11wne_d_TB}$c$) enables the TB to overcome this wall-ward attraction and migrate to the unbounded side of the LB's wake by $t \approx 12$. This evolution in the TB's position relative to the LB's wake is highlighted in figure \ref{fig:11wne_d_TB}$(d)$, where the structure of the spanwise vorticity in the symmetry plane $Z=0$ is shown at $t = 6$ (left) and 12 (right). Once the TB reaches the unbounded side of the LB, it begins to escape, as the shear-induced lift now points away from the wall. 
These evolutions are similar to those in the unsteady WNE scenario described in the previous section, except that the strong interplay between the shed vortices and the TB no longer occurs. The absence of this strong vortical interaction is confirmed by examining the evolution of the streamwise vorticity field during the TB's escape (without figure). It turns out that while there is still shedding of streamwise vortices from the LB's surface as it migrates away from the wall, the intensity of the shed vortices is relatively weak; this is primarily the reason the LB cannot undergo a periodic near-wall bouncing motion\citep{2024_Shi_b}. Owing to the absence of this strong vortical interaction, the maximum lateral velocity of the TB during the escape is only $70\%$ of that in the unbounded condition (0.39 over 0.56), which is smaller than the ratio of $86\%$ (0.68 over 0.79) achieved in the unsteady WNE scenario depicted in figure \ref{fig:7wne_b_path}$(c)$.

\begin{figure}
    \centering
    \includegraphics[height=8cm]{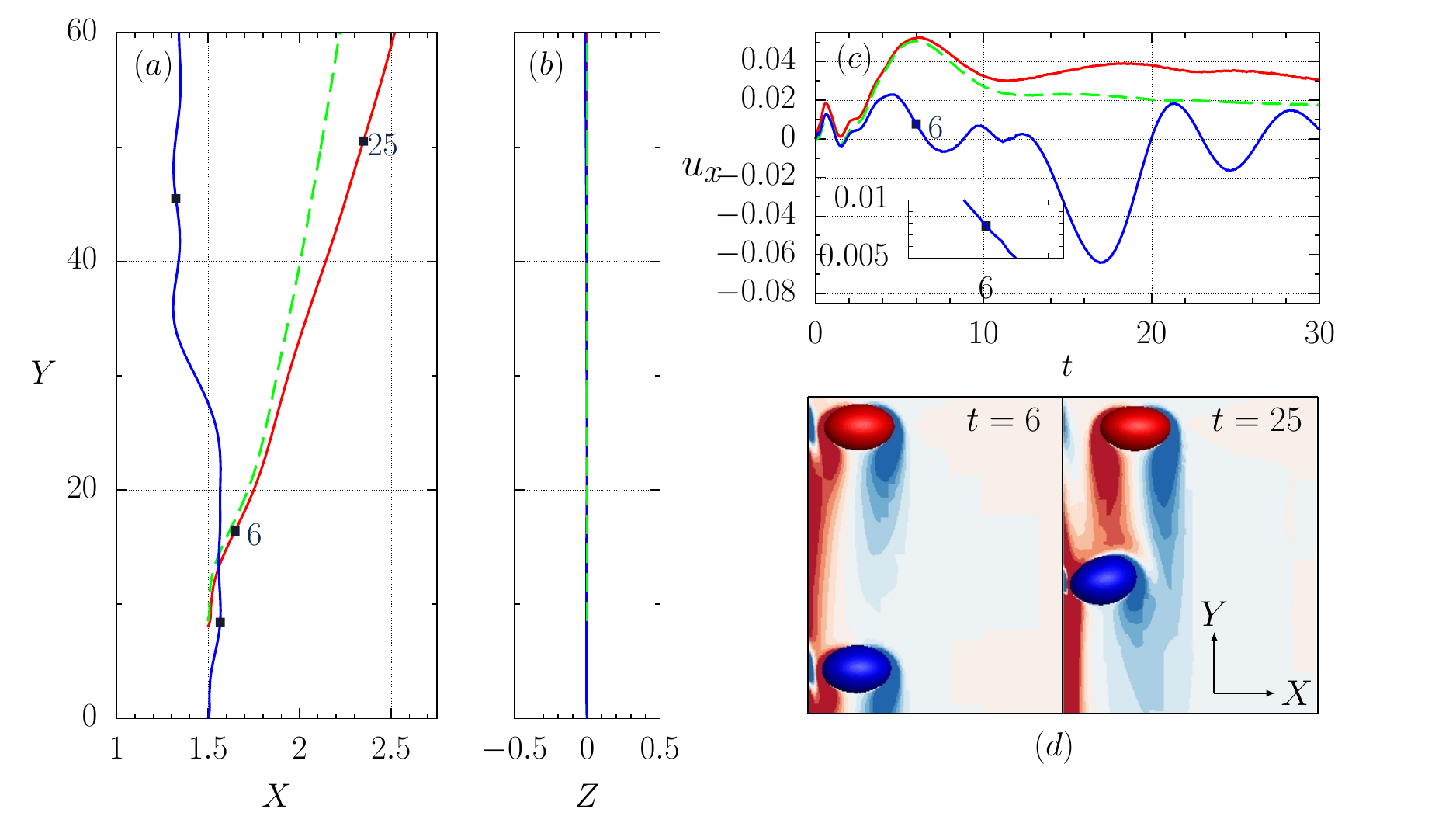}
    \caption{Same as figure \ref{fig:11wne_d_TB} but in a quasi-steady, LB-led WNE scenario $(\Ga, \Bo) = (20, 0.5)$. At $t=6$, The wall-normal velocity of the TB is about 0.008.}
    \label{fig:12wne_d_LB}
\end{figure}

One of the key elements for the TB to escape is its initial departing velocity, denoted as $u_d^\text{TB}$, where $u_d^\text{TB} \approx u_x^\text{TB}$ when $Y^\text{TB} = \Delta_y^0$. This velocity must be sufficiently \q{large} to overcome the wall-ward attraction induced by the LB's wake. While $u_d^\text{TB}$ results from the vortical wall-bubble interaction, its magnitude largely depends on the closest wall distance the TB achieves before this vortical interaction commences. As this phase lasts for a duration of about $\Ga(X^0)^2$, $u_d^\text{TB}$ decreases with decreasing $\Ga$ and, at very low $\Ga$, may not be large enough for the TB to overcome the wall-ward attraction. An instance of this event is observed in the case $(\Ga, \Bo) = (20, 0.5)$. As illustrated in figure \ref{fig:12wne_d_LB}$(a)$, the TB reaches the LB's initial vertical position at $t \approx 6$, where $u_d^\text{TB}$ is about 0.008 (see $u_x^\text{TB}$ at $t = 6$ in figure \ref{fig:12wne_d_LB}$c$), which is almost seven times smaller than that for $(\Ga, \Bo) = (30, 0.5)$. Given the small $u_d^\text{TB}$, the TB briefly migrates away then stabilizes at a wall-normal position from $t \approx 8$ to 12. During this time period, the LB moves away from the wall at a larger velocity, increasing the horizontal separation over time. From $t \approx 12$, the pair interaction becomes strong enough for $u_x^\text{LB}$ to exceed that in the LB-only condition. Consequently, while the TB remains trapped in the gap between the LB and the wall, the two bubbles gradually move apart in the wall-normal plane. By $t = 25$, the horizontal separation exceeds $2$, indicating that the TB is no longer within the wake of the LB. 

\begin{figure}
    \centering
    \includegraphics[height=8cm]{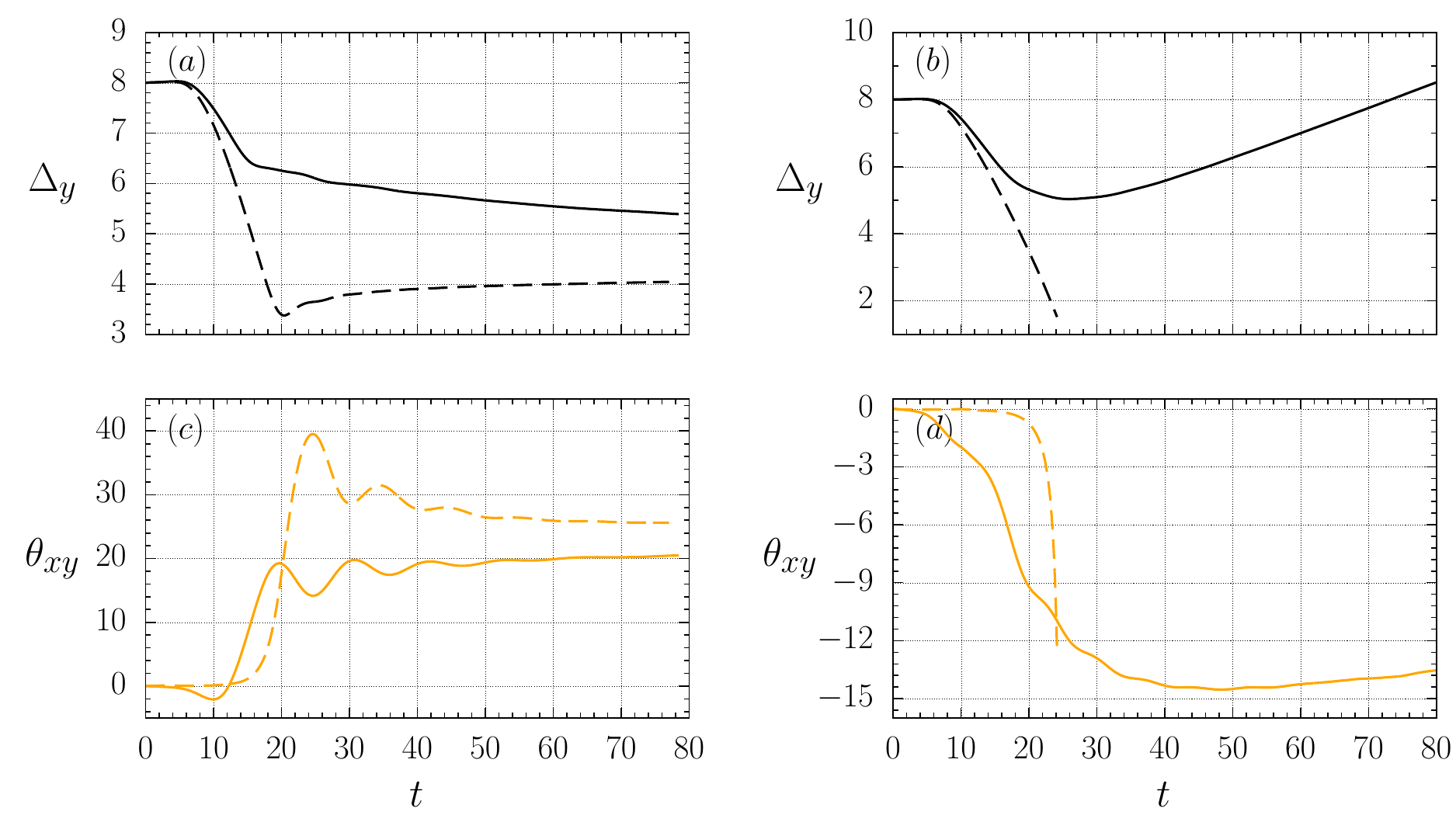}
    \caption{Results for the vertical separation and the inclination angle obtained for  $(\Ga, \Bo) = (30, 0.5)$ (left column) and $(20, 0.5)$ (right column). The solid and dashed lines represent results in the wall-bounded and the unbounded conditions, respectively.}
    \label{fig:13wne_d_geo}
\end{figure}

To understand the effects of the wall on the evolution of the geometry of the bubble pair, we present in figure \ref{fig:13wne_d_geo} results for the vertical separation and the inclination angle for the two cases mentioned above. In both cases, the wall facilitates the escape, aligning with observations in the unsteady WNE scenario. Specifically, for the case with $(\Ga, \Bo) = (20, 0.5)$, the evolution in the unbounded condition indicates that the two bubbles coalesce through a head-on approach, i.e., without any prior escape of the TB. This coalescence is absent in the wall-bounded condition, as the wall encourages the bubble pair to separate in the wall-normal direction, thus avoiding the head-on coalescence. Moreover, unlike in the (unsteady) TB-led WNE scenario where the wall retarding effect on the LB promotes the TB to finally catch up, in a LB-led WNE scenario, the wall retarding effect is more pronounced on the TB, thus inhibiting its catch-up (see $\Delta_y$ in figure \ref{fig:13wne_d_geo}$b$).

\subsection{Influence of initial angular deviations} \label{sec:3-4par_eff}
For the discussions above, the results were obtained under the inline configuration where, at the initial state, the wall-normal and wall-parallel positions of the two bubbles are identical. In practice, especially under experimental conditions, bubble pairs might not always stay initially inline, leading to sizable angular deviations of the bubble pair prior to the pair or wall-bubble interaction. The resulting influence on the evolution of the bubble path will be addressed in this section. For this purpose, we have carried out two series of additional runs within the same range of $\Ga$ and $\Bo$ as considered in the inline configuration, but with small yet finite initial angular deviations in either the wall-parallel or the wall-normal planes. In both series of runs, we have kept the initial vertical separation between the two bubbles. Regarding the initial horizontal positions, the position of the LB is kept unchanged, while that of the TB is adjusted to either $(1.5, -\Delta_z^0)$ or $(1.5 - \Delta_x^0, 0)$. This results in initial angular deviations of $\theta_{zy}^0 = \tan^{-1}(\Delta_z^0 / 8)$ in the wall-parallel plane and $\theta_{xy}^0 = \tan^{-1}(\Delta_x^0 / 8)$ in the wall-normal plane. Recall that a positive $\Delta_x$ indicates that the LB is located at a larger coordinate value along that the $X$ dimension. Hence, a positive $\theta^0_{xy}$ means the TB is initially closer to the wall than the LB.

The sign and magnitude of $\theta_{zy}^0$ and $\theta_{xy}^0$ deserve further discussion. At the initial state, the problem in the inline configuration is symmetric with respect to $Z=0$. With an initial angular deviation in the wall-parallel plane, the sign of $\theta_{zy}^0$, which specifies the angular deviation with respect to $Z=0$, is irrelevant. Following \citet{2021_zhang}, we set $\theta_{zy}^0=-1.8^\circ$, corresponding to $-\Delta_z^0=0.25$, which is an order of magnitude larger than the finest grid cells. The situation becomes different with an initial angular deviation in the wall-normal plane. Given the presence of the wall, the inline problem is not symmetric with respect to $X=0$. With an initial wall-normal deviation, the sign of $\theta_{xy}^0$ specifies whether the TB is closer to or farther away from the wall and will definitely play a role. In practice, given the wall-induced retarding effect, a trailing bubble can more easily catch up and interact with the wake of a leading bubble if it stays farther away from the wall. Hence, $\theta_{xy}^0$ arises from two causes: one from the unavoidable disturbance in the ambient flow, which is also the key cause of $\theta_{zy}^0$, and the other from the wall retarding effect, which is irrelevant in causing $\theta_{zy}^0$. Considering these factors, it seems more reasonable to consider a \emph{negative} and relatively larger angular deviation in the wall-normal plane. Following this spirit, we set $-\Delta_x^0=0.5$, corresponding to a deviation angle of $\theta_{xy}^0=-3.6^\circ$.

\begin{figure}
    \centering
    \includegraphics[height=8.15cm]{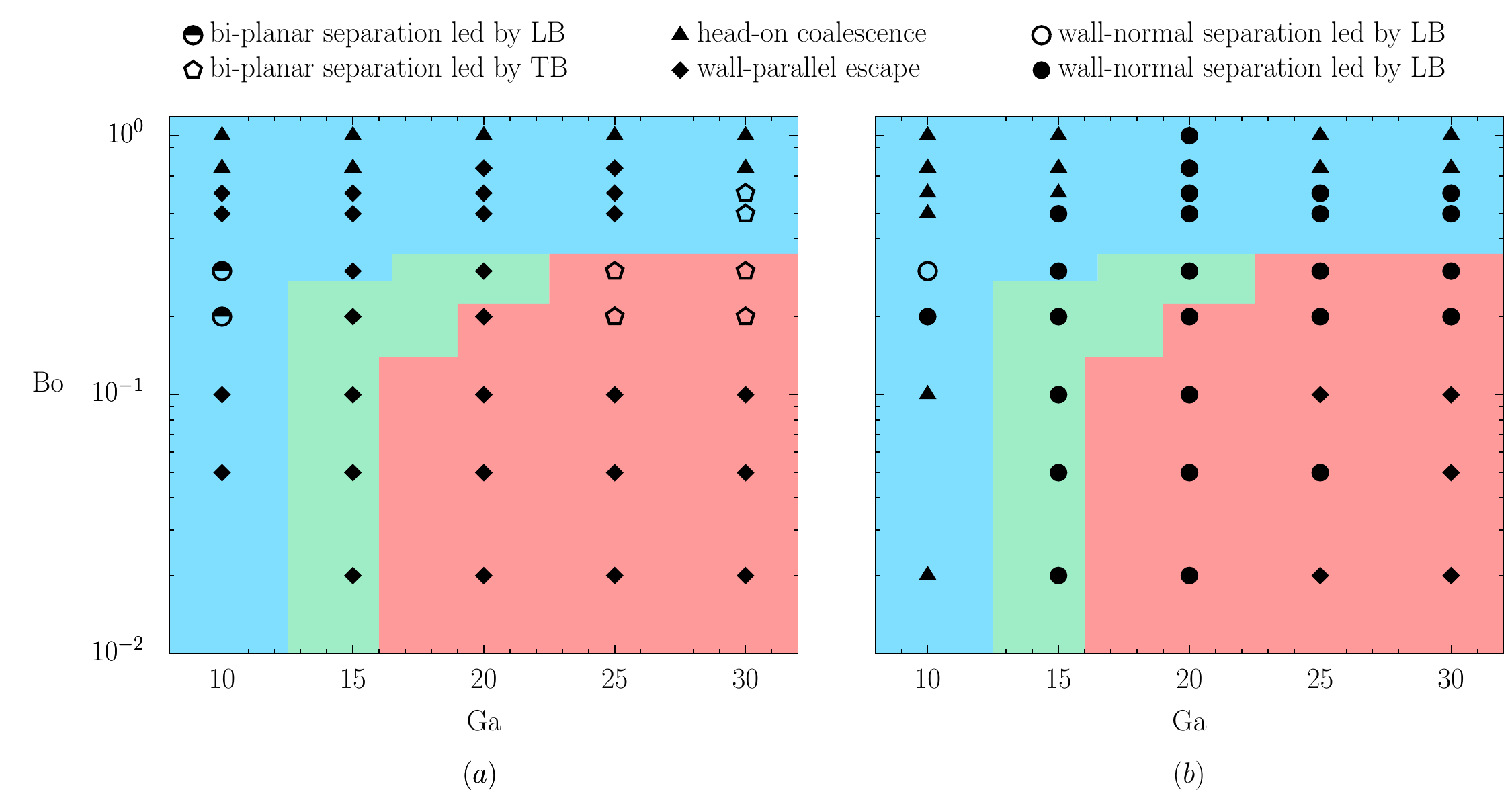}
    \caption{Phase diagram in the $(\Ga,\Bo)$ plane showing the resulting path of a bubble pair released with an angular deviation of ($a$) $1.8^\circ$ in the wall-parallel plane (i.e., $\theta_{zy}^0=-1.8^\circ$) and ($b$) $3.6^\circ$ in the wall-normal plane (i.e., $\theta_{xy}^0=-3.6^\circ$). Following figure \ref{fig:2phase}, the shape of the symbols and the background colour are used to distinguish the type of motion in the pair configuration and in the corresponding single-bubble configuration, respectively. The symbols correspond as in figure \ref{fig:2phase}. Note the first two new symbols labeled above denote bi-planar separation scenario; $\bilb$ represents bi-planar separation in which the separation on wall-normal plane is led by LB, whereas $\bitb$ represents those led by TB.}
    \label{fig:14phase_angular}
\end{figure}

\begin{figure}
    \centering
    \includegraphics[height=8cm]{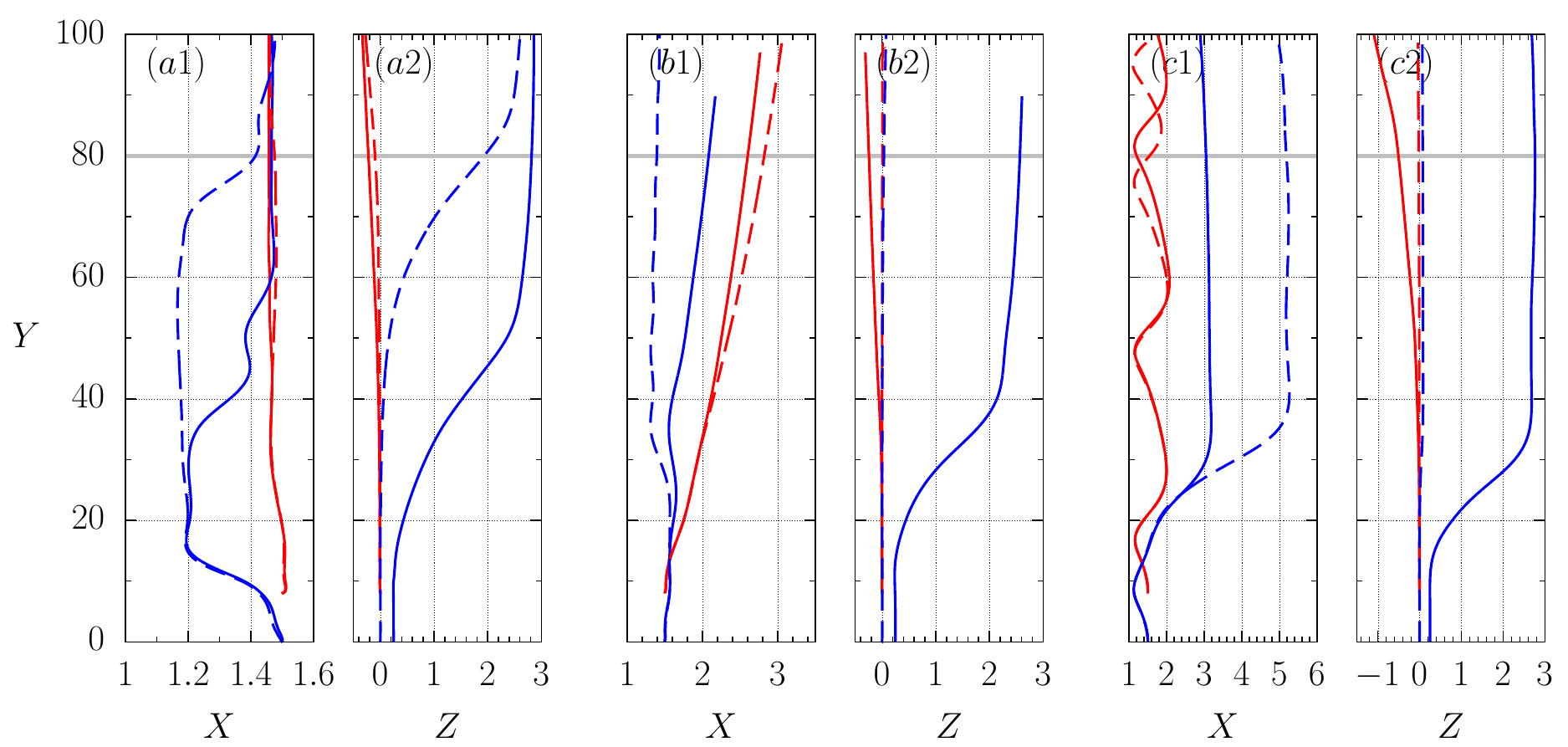}
    \caption{Influence of an initial, wall-parallel angular deviation on the path of the bubble pair. Dashed lines: $\theta_{zy}^0=0$; solid lines: $\theta_{zy}^0=-1.8^\circ$. From left to right, the three pairs of panels correspond to cases with $(\Ga, \Bo) = (15, 0.2)$, $(20, 0.5)$, and $(25, 0.2)$. The horizontal gray line indicates the height of $Y=80$.}
    \label{fig:15Z0_path}
\end{figure}

Figure \ref{fig:14phase_angular} summarizes the resulting path with two different initial angular deviations as outlined above for $10\leq \Ga\leq30$ and $0.02\leq \Bo\leq 1$. In addition to the various types of path evolutions identified in the inline configuration (see figure \ref{fig:2phase}$a$), a new type of motion arises when considering the initial angular deviation in the wall-parallel plane (see new symbols in figure \ref{fig:14phase_angular}$a$). In this scenario, the path separation occurs in both the wall-normal and the wall-parallel planes (see figure \ref{fig:15Z0_path}($c1$, $c2$) for the path). We identify such a bi-planar separation scenario based on the following criteria: In the quasi-steady flow regime where the bubble in the LB-only condition either stabilizes at a wall-normal position close to the wall or migrates away from the wall, bi-planar separation is identified if the horizontal separations in both planes exceed $1.5$. Conversely, in the unsteady regime where the bubble repeatedly bounces close to the wall in the LB-only condition, a separation in the wall-normal plane is confirmed only if the TB manages to escape and catch up with the LB.

Comparing the two phase diagrams in \ref{fig:14phase_angular} with that in figure \ref{fig:2phase}($a$), it may be concluded that in general, an initial angular deviation promotes the path separation in the same plane. Specifically, for an angular deviation in the wall-normal plane with TB placed initially farther away from the wall, the path separation is governed by an escape led by the TB, except on two occasions. When $\Ga$ is small ($\Ga \leq 15$), this angular deviation promotes coalescence: it lowers the critical $\Bo$ beyond which a head-on coalescence occurs and also stimulates the coalescence of the bubble pairs at low $\Bo$. This promotion may be understood by noting that the larger initial wall distance permits the TB to rise faster, thus promoting the head-on collision. The other exceptional occasion specifically concerns the case where $(\Ga, \Bo) = (30, 0.1)$. In this case, the path transitions from a wall-normal escape to a wall-parallel escape, both led by the TB. This transition yields a suppression of the wall-normal separation by a wall-normal angular deviation.
To understand this \q{abnormal} response of the bubble path, the corresponding result is examined (without figures). It turns out that the path evolution is similar to that in an unsteady wall-parallel escape scenario (see figure \ref{fig:5wpe_b_path}). More specifically, owing to the increased wall-normal separation of the TB, the lateral velocity of the TB remains negative (i.e., towards the wall) when the TB first meets the wake of the LB. This latter occasion highlights the difficulty in precisely qualifying the influence of a wall-normal angular deviation in the unsteady flow regime. In this regime, whether a wall-normal angular deviation would lead to a separation in the same plane depends on the lateral position and velocity of the TB when it rises to the LB's initial position. Given the oscillatory nature of the lateral motion, both values do not vary monotonously with the angular deviation even with a fixed initial vertical separation. In the following, we discuss the results of several selected cases that help illustrate in more detail the influence of the angular deviations.

Figure \ref{fig:15Z0_path} compares the paths obtained with $\theta_{zy}^0=-1.8^\circ$ to their counterparts in the inline configuration. From left to right, the three pairs of panels correspond to cases with $(\Ga, \Bo) = (15, 0.2), (20, 0.5)$, and (25, 0.2). In the inline configuration, the corresponding path evolutions follow the scenarios of the WPE, LB-led WNE, and TB-led WNE, respectively. In the presence of a wall-parallel angular deviation, the separation in this plane is promoted, in line with that observed in the unbounded configuration \citep{2021_zhang}. Specifically, in the case where the path follows a WPE scenario in the inline configuration [panels $(a1, a2)$], the TB begins its lateral drift after a vertical displacement of about $20$, which is only half that observed in the inline configuration. In the other two cases where the path follows a WNE scenario, this angular deviation strongly inhibits separation in the wall-normal plane. Consider, for instance, the case with $(\Ga, \Bo) = (20, 0.5)$ [panels $(b1, b2)$], where the wall-normal separation is led by the LB. Here, the wall-normal separation remains smaller than $0.5$ when the vertical displacement of the TB reaches $80$, whereas the corresponding value is about $1.5$ in the inline configuration. For the other WNE case, where the separation is led by the TB [panels $(c1, c2)$], the small initial angular deviation reduces the maximum wall-normal drift of the TB by about 40\%.

\begin{figure}
    \centering
    \includegraphics[height=8cm]{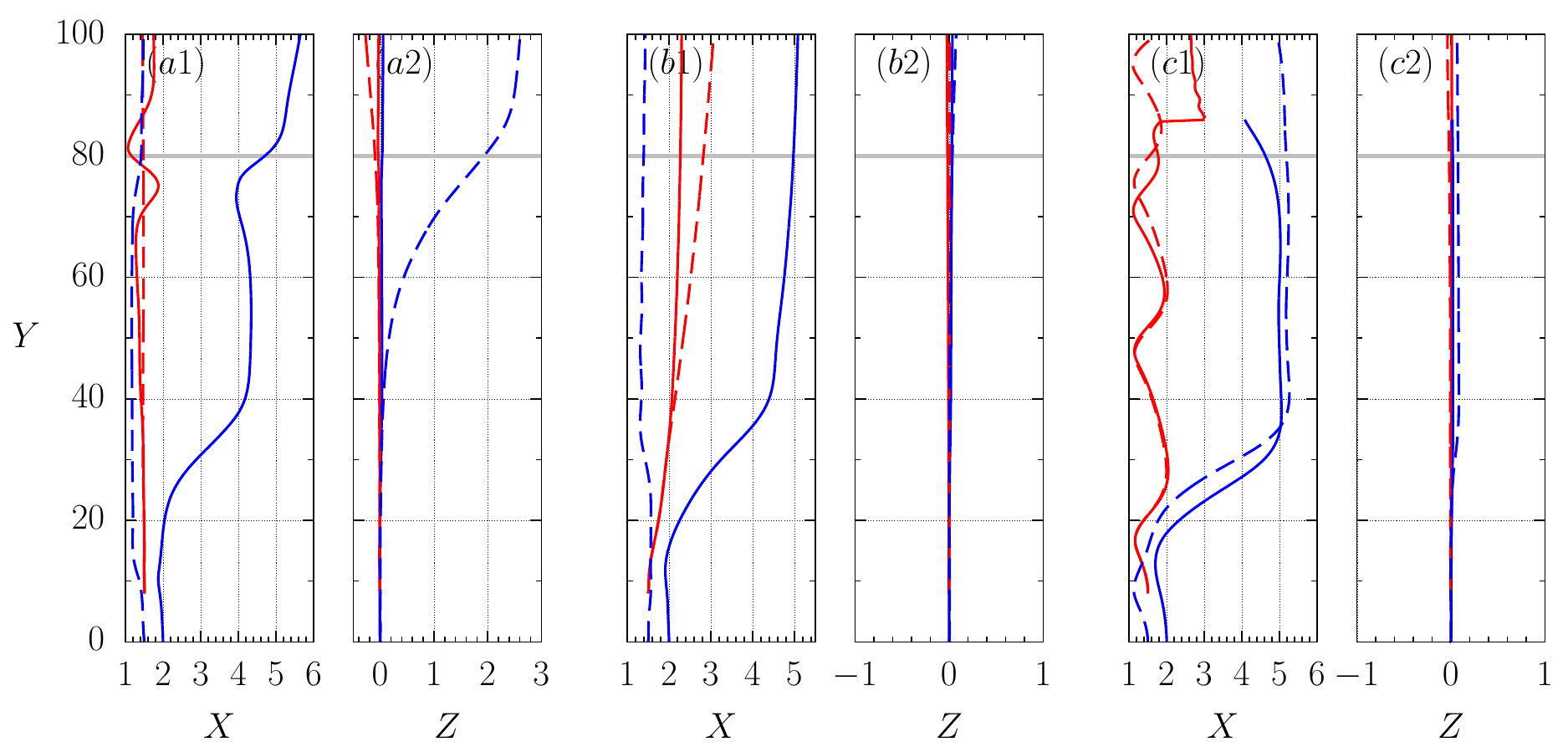}
    \caption{Same as figure \ref{fig:15Z0_path} but with solid lines denoting results with $\theta_{xy}^0=-3.6^\circ$.}
    \label{fig:16X0_path}
\end{figure}

Figure \ref{fig:16X0_path} provides a similar comparison but now focuses on an angular deviation in the wall-normal plane. In all three cases with an initial angular deviation, the TB always stays on the unbounded side of the LB's wake when it reaches the initial position of the LB. Hence, while the wall-normal velocity remains negative at this moment, the shear-induced lift from the LB's wake is capable of driving the TB away from the wall. For a bubble pair following a WPE scenario in the inline configuration [panels $(a1, a2)$], this angular deviation inhibits the separation in the wall-parallel plane, leading to a transition from a WPE scenario to a TB-led WNE scenario. For a bubble pair following a LB-led, WNE scenario in the inline configuration [panels $(b1, b2)$], this angular deviation suppresses the wall-normal departure of the LB such that the bubble leading the separation is switched from the LB to the TB. Finally, for a bubble pair following a TB-led, WNE scenario [panels $(c1, c2)$], the angular deviation leads to an earlier escape, as the TB no longer needs to overcome the wall-ward attraction induced by the asymmetric wake of the LB (see, for instance, the discussion in \S~\ref{sec:3-3-1wne_b}). This is why, once the escape is completed, the wall-normal distance of the TB is smaller than that in the inline configuration. 

\subsection{Implications for the formation of near-wall horizontal bubble clusters}\label{sec:3-5mac}
Clustering of bubbles in wall-bounded bubble-laden flows has been frequently reported in prior works \citep{mathai2020bubbly}. Several distinct mechanisms are known to play a role. In addition to the wall-bubble and pairwise interactions outlined in this work, of particular note are the shear-induced lift due to velocity gradients in the background mean flow \citep{1987_Auton, 1998_Legendre, 2002_Tomiyama}, the turbophoresis effect caused by gradients in liquid velocity fluctuations \citep{reeks1983transport, 2015_Santarelli, santarelli2016direct}, and turbulent diffusion due to bubble volume fraction gradients \citep{balachandar2010turbulent}. Clarifying how these individual mechanisms combine to alter the structure of the cluster is beyond the scope of the present work. Instead, we restrict our attention to clustering processes in upward channel flows at moderate channel Reynolds numbers. In such a configuration, while turbulence-related effects are subtle, wall-bubble and pairwise interactions play a significant role, typically leading to the formation of near-wall horizontal bubble clusters (i.e., bubbles aligning horizontally within the wall-parallel plane). This is observed, for instance, in experiments with approximately 1 mm diameter bubbles rising in water \citep{2008_Takagi, 2021_Maeda}, corresponding to $(\Ga, \Bo) = (35, 0.03)$, and in the numerical simulation by \citet{2013_Lu}, where a case with $(\Ga, \Bo) = (43, 0.125)$ was considered. In these studies, the ambient flow is not stationary but moves upwards. Due to the ambient shear, all bubbles experience a lateral shear-induced lift directed toward the wall. This promotes the formation of a bubble-rich wall layer at a position where the shear-induced lift towards the wall balances the inertial wall-induced lift pointing away from the wall \citep{2020_Shi}.

Nevertheless, the formation of such a bubble-rich wall layer does not necessarily mean that bubbles will cluster horizontally. Before addressing the various clustering mechanisms identified in prior work, it is necessary to consider the influence of the ambient shear, which is not accounted for in these mechanisms. In the upward channel flows considered in \cite{2008_Fukuta} and \cite{2021_Maeda}, the shear-induced lift consistently points towards the wall. Therefore, it does not directly affect the separation of the bubble pair in the wall-parallel plane, as the shear rate is uniform within this plane. However, it 'suppresses' the separation of the bubble pair in the wall-normal plane. This occurs because the shear-induced lift brings the equilibrium position closer to the wall. As the repulsive wall force varies more rapidly with wall distance at smaller separation \citep{2024_Shi_a}, this shift towards the wall increases the ‘stiffness’ of the wall-bubble system at the equilibrium point, thereby suppressing the onset of separation in the wall-normal plane. Given this suppression, one may expect the ambient shear to promote the separation in the wall-parallel plane. Nevertheless, whether this promoted separation will ultimately lead to the formation of horizontal bubble clusters depends largely on the final geometrical configuration formed by the bubble pair, particularly on whether the bubble pair may rise side-by-side or not.

Previous works on pair bubble interactions have indicated the existence of two mechanisms promoting the side-by-side rising of bubble pairs, and hence, the horizontal clustering \citep{1993_Kok_a, 1993_Kok_b, 1994_Yuan, 1997_Harper, 2003_Legendre, 2011_Hallez, 2021_zhang}. The first is an irrotational mechanism derived from potential flow theory. In this context, two spherical bubbles that are vertically aligned initially repel each other, while rotating around each other such that their line of centres becomes horizontally aligned, after which they approach each other along a horizontal line until they touch (see \citet{2021_Maeda}, for example). The second is the DKT mechanism, already introduced in \S~\ref{sec:1intro}. This mechanism manifests for bubbles of $\Rey = O(10)$ \citep{2019_Kusuno,2021_zhang} and for droplets with a fluid-to-drop viscosity ratio of $O(1)$ at $\Rey = O(100)$ \citep{2016_Bayareh}. This mechanism primarily involves an attractive wake effect, whereby two particles or bubbles initially aligned vertically are first attracted toward each other, then repel in the horizontal direction until they reach an equilibrium separation and fall/rise side by side. 

Unfortunately, both type of pair interaction was not observed in our three-dimensional simulations carried out under wall-bounded conditions.
Specifically, the results from an additional series of run carried out at $(\Ga, \Bo) = (30, 0.05)$ (see appendix \ref{sec:appA1}) reveal that the path of a bubble pair at comparable $\Rey$ and $\upchi$ with that of a $1~\text{mm}$ diameter bubbles rising in water follows the WPE scenario. The geometry of the bubble pair obtained in the final state yields final inclinations in the range $15^\circ-30^\circ$, wall-normal separations ranging from 4 to 5 radii, and vertical separations from 8 to 14 radii, depending on the initial angular deviation in the wall-parallel plane. Given the unavoidable variations of initial angular deviations in real bubbly flows, this relatively wide range of near-equilibrium inclinations and separations corresponds to a very homogeneous spatial bubble distribution within the bubble-rich wall layer.

Nevertheless, we strongly anticipate that it is the DKT mechanism that is responsible for the formation of the horizontal clusters observed in the previous investigations. To clarify this, note that in the numerical simulation by \citet{2013_Lu}, the gas-to-liquid (dynamic) viscosity ratio considered is $\mu^\ast=1$, indicating that the particles under consideration were viscous droplets rather than inviscid bubbles, which correspond to $\mu^\ast\ll1$. Given this difference, the amount of vorticity produced at the \q{bubble} surface, and hence the vortical pair interaction, are considerably higher \citep{2001_Feng, 2020_Rachih, 2024_Shi_c} than that for clean bubbles. This is known \citep{2021_zhang} to enhance the attractive effect of the LB's wake, promoting the DKT scenario. On the other hand, in the relevant experimental investigations, a certain amount of surfactants were used to prevent coalescence. This is known to lead to an increase in drag. Specifically, the measured Reynolds number based on the mean near-wall rising velocity of an isolated bubble is 205 \citep{2021_Maeda}, about 20\% lower than that in the corresponding clean gas-liquid system \citep{2024_Shi_b}. While this level of contamination is still incapable of reversing the sign of the shear-induced lift \citep{2008_Fukuta}, it may already cause a sizable decrease in the pressure at the rear part of the bubble \citep{1997_Cuenot, 2023_Kentheswaran, 2024_Rubio}, and hence an enhanced attractive effect of the LB's wake which promotes the DKT scenario \citep{1993_Kok_b, 1998_Duineveld, 2016_Kusuno}.

Finally, it must be acknowledged that confirming whether the DKT mechanism actually manifests in real bubble-laden channel flows requires detailed information on the time evolution of the horizontal clustering. Unfortunately, this information is not available from the previous investigations.

\section{Summary and concluding remarks}\label{sec:4conclude}

We carried out three-dimensional simulations for a pair of deformable bubbles rising initially inline and close to a vertical wall in an otherwise quiescent liquid. The range of parameters considered is $10 \leq \Ga \leq 30$ and $0.02 \leq \Bo \leq 1$. In the absence of the wall, the bubble pair undergoes lateral separation, following either a DKT or a side-escape scenario \citep{2021_zhang}, provided that $\Bo$ is smaller than a critical $\Ga$-dependent value $\Bo_{c1}(\Ga)$. For $\Bo > \Bo_{c1}(\Ga)$, the unbounded bubble pair coalesces through a head-on approach.

The presence of the wall makes the undisturbed flow field non-axisymmetric, although the bubble pair is initially rising inline. Specifically, the wall elongates the wake of the LB on the wall-facing side (see the vortical structure in figure \ref{fig:3wpe_r_path}$a$). Once this elongated wake has been advected downstream over a distance comparable to the initial vertical separation, it leads to an attractive shear-induced lift on the TB, causing the sideways forces experienced by the two bubbles to differ. This wall-induced asymmetry drastically changes the evolution of the bubble paths. In the high-$\Bo$ regime, head-on coalescence is avoided up to a larger $\Bo_{c1}(\Ga)$. For $\Bo < \Bo_{c1}(\Ga)$, while the bubble pair always separates laterally, the DKT scenario, which occurs in the unbounded condition for $\Ga \lesssim 12$, is absent. The horizontal separation occurs in either the wall-normal plane or the wall-parallel plane, depending on the competition between the irrotational and the vortical effects:
\begin{itemize}
\item When $\Bo$ is below a critical $\Ga$-dependent threshold $\Bo_{c2}(\Ga)$, the irrotational effect dominates, causing the two bubbles to initially attach to the wall. Due to the asymmetric wake of the LB, the TB rests at a (mean) position closer to the wall, while the LB either remains near the wall or oscillates laterally. This two-dimensional configuration persists until the difference in the rising speed of the two bubbles reaches a critical, case-dependent value. Afterwards, the bubble pair separates in the wall-parallel plane, similar to a side-escape scenario in the unbounded condition. That is, the TB escapes laterally from the wake of the LB without significantly altering the path of the latter.\par
\item For $\Bo_{c2}(\Ga) < \Bo < \Bo_{c1}(\Ga)$, the vortical effect dominates, causing both bubbles to migrate away from the wall after a brief irrotational interaction with the wall. Due to the wall-ward attraction by the asymmetric wake of the LB, the TB usually departs at a slower speed than the LB. This difference makes the inclination angle in the wall-normal plane ($\theta_{xy}$) to increase in time. Once $\theta_{xy}$ reaches sizable values, the horizontal separation of the bubble pair is further accelerated by the repulsive pair interaction. Consequently, while the TB remains trapped between the wall and the LB, it can \q{escape} from the LB's wake due to the faster departure of the latter.\par
\item In both regimes, there is an exceptional scenario where the separation takes place in the wall-normal plane, as in the second regime, but the separation is led by the TB, not the LB. In this scenario, the TB gains a finite departing velocity prior to the pair interaction, enabling it to overcome the wall-ward attraction and reach the unbounded side of the LB's wake. Once this process is completed, the shear-induced lift by the LB's wake becomes repulsive, causing the TB to migrate away from the wall at a greater velocity and hence an escape in the wall-normal plane led by the TB. In such escapes taking place in the unsteady flow regime where streamwise vortex shedding is pronounced, we also identified a distinct vortical interaction between the shed vortices (from the LB) and the TB (figure \ref{fig:9wne_b_vor}). Such vortical interaction leads to an enhancement in the streamwise vorticity in the wake of the TB, making its lateral drift to exceed that in the unbounded condition (see figure \ref{fig:7wne_b_path}$a$).
\end{itemize}

The presence of the wall also influences the final geometry of the bubble pair. This influence can be examined by comparing the final geometry formed in the two TB-led escape scenarios—with the separation either in the wall-normal or the wall-parallel plane—with those in the corresponding unbounded configuration. This final geometry may be characterized using the vertical separation $\Delta_y$ and the inclination angle, either $\theta_{xy}$ or $\theta_{zy}$, depending on the plane of separation. \\
- In cases where the bubble pair separates in the wall-parallel plane, provided that the flow remains in the quasi-steady regime such that no strong vortex shedding takes place, the final $\Delta_y$ (respectively, $\theta_{zy}$) is larger (respectively, smaller) in the presence of the wall (see figure \ref{fig:3wpe_r_path}). These modifications are sustained but less significant when the flow is in the unsteady flow regime (see figure \ref{fig:6wpe_b_velo}), where both bubbles bounce repeatedly close to the wall. These features may be understood by noting that the presence of the wall leads to a faster decay of the disturbance, thereby attenuating the wake effect of the LB. This attenuation is less significant when the bubble starts to bounce close to the wall, making the final $\theta_{xy}$ and $\theta_{zy}$ align closely with those in the unbounded condition.\\
- For the TB-led escape scenarios taking place in the wall-normal plane, the influence of the wall on the final geometry largely depends on the dominant mechanism for the wall-bubble interaction. Provided that the vortical effect dominates, both bubbles migrate away from the wall. In this context, the influence of the wall is similar to that in a wall-parallel escape scenario. Specifically, the wall leads to an early separation (with respect to that in the unbounded condition), resulting in, in the terminal state, a larger $\Delta_y$ but a smaller $\theta_{xy}$ (see figure \ref{fig:13wne_d_geo}). Things become drastically different when the wall-bubble interaction is governed by the irrotational effect. In this context, the LB remains close to the wall during the TB's escape. Given the no-slip condition at the wall, the LB also experiences a retarding effect from the wall which promotes the TB catching up (see figures \ref{fig:8wne_b_geo} and \ref{fig:10wne_b_nocoal}). This catch-up event is absent in the corresponding unbounded condition \citep{2021_zhang}. A consequence of this catch-up event is that, provided the bubble pair does not coalesce, the pair interaction may drive the bubble pair to migrate away from the wall while rising side-by-side (see figure \ref{fig:10wne_b_nocoal}). This escape from the wall may be closely linked to the formation of horizontal bubble clusters in the core region of the pipe flow \citep{2013_Lu, 2015_Santarelli}. 

Finally, to assess the generality of the regime map (figure \ref{fig:2phase}) obtained with fixed initial wall-normal distance of the bubble pair ($X^0$) and their vertical separation ($\Delta_y^0$), parametric studies in typical configurations have been carried out.

- In the first series of tests, we considered a misaligned configuration, with the level of misalignment characterized by the initial angular deviations. Two additional series of runs within the same range of $\Ga$ and $\Bo$ as considered in the inline configuration, considering an initial angular deviation in either the wall-parallel or the wall-normal planes, respectively. It turns out that the influence of the angular deviation depends on the plane in which the initial offset occurs. With an angular deviation in the wall-parallel plane (figures \ref{fig:14phase_angular}($a$) and \ref{fig:15Z0_path}), the path separation in this plane is always promoted. Specifically, in cases where separation occurs in the wall-normal plane in the inline configuration, the path may now follow a bi-planar separation [see figure \ref{fig:15Z0_path}$(c1, c2)$], one in the wall-normal plane as in the inline configuration, and the other in the wall-parallel plane due to the initial offset. The influence becomes different with an angular deviation in the wall-normal plane (figures \ref{fig:14phase_angular}$(b)$ and \ref{fig:16X0_path}). This difference primarily arises in the unsteady flow regime, where the bubble starts to bounce in the LB-only condition. In this regime, whether a wall-normal angular deviation leads to a wall-normal separation depends on the lateral position and velocity of the TB when it rises to the LB's initial position. Given the oscillatory nature of the lateral motion, both values do not vary monotonously with the angular deviation, even with a fixed initial vertical separation.

- The second series of runs focuses on the inline configuration, but with varying $X^0$ and $\Delta_y^0$ (see Appendix \ref{sec:appA2}). These new results provide further insights into how the regime map changes with $X^0$ and $\Delta_y^0$ (see the discussion therein for details). Based on these new findings and those addressing the influence of initial angular deviations, it may be concluded that the regime map shown in figure \ref{fig:2phase} may still serve as a reference for different $(X^0,\Delta_y^0)$ values, provided that the case specified by $(\Ga, \Bo)$ lies in the steady flow regime where no vortex shedding occurs. However, the fate of the bubble pair may change drastically with variations in $(X^0,\Delta_y^0)$ if the case lies in the unsteady flow regime, owing to the oscillatory nature of the lateral motion as mentioned above.

There are two key aspects that should be addressed in future work. While we have already explored the influence of pairwise misplacement in this study, the assumption of identical bubbles in the simulations remains a limitation. Thus the first aspect is to further investigate the differences between the bubbles, an area of research that is currently underway. This includes examining variations in bubble size which leads to differences in physical parameters such as $\Ga$ and $\Bo$. These differences, similar to the misplacement of bubbles, are commonly encountered in real experimental settings. The second aspect is the development of a force-balanced model, akin to those proposed for unbounded situations in \cite{2011_Hallez,2021_Maeda}. However, the presence of a wall and the slight deformation of the bubbles add complexity, making the formulation of such a model more challenging. Addressing these two aspects will be the focus of future research, extending beyond the scope of the current study.

\appendix
\section{Influence of spatial configuration}
\label{sec:appA}
In this appendix, we present three sets of simulations to support our conclusions from the main text.
\subsection{Influence of wall-parallel angular deviation for $(\Ga, \Bo) = (30, 0.05)$}
\label{sec:appA1}
\begin{figure}
    \centering
    \includegraphics[height=11cm]{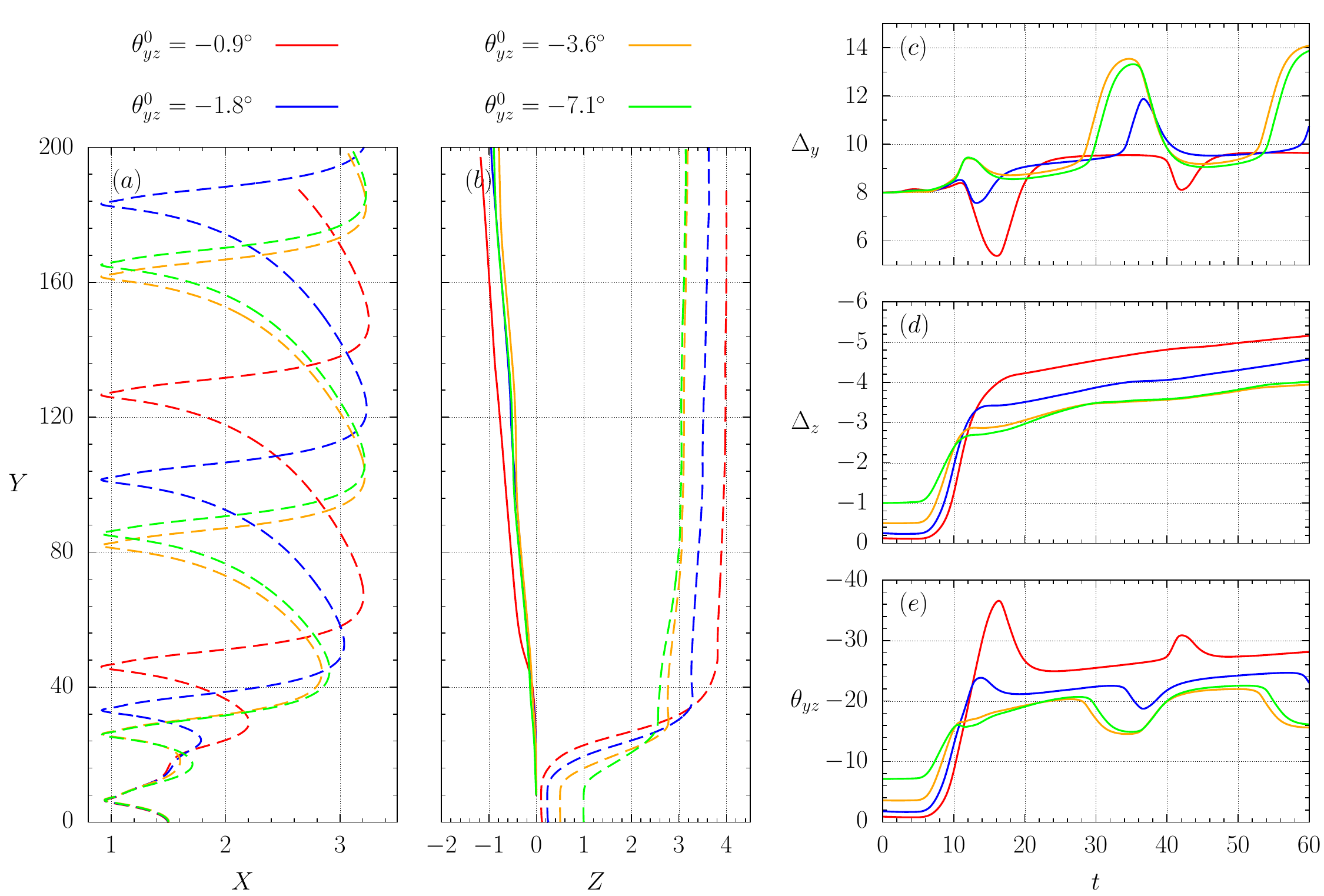}
    \caption{Results for selected motion characteristics for the case with $(\Ga,\Bo) = (30,0.05)$, obtained with different wall-parallel angular deviations. $(a,b)$: Trajectories in the $(X,Y)$ and $(Z,Y)$ planes, respectively, where solid and dashed lines represent LB and TB, respectively. $(c-e)$: Time evolutions of $\Delta_y$, $\Delta_z$, and $\theta_{zy}$, respectively. Different colors denote different $\theta_{zy}^0$ as indicated in the figure.}
    \label{fig:17appen}
\end{figure}

As a supplement to \S~\ref{sec:3-4par_eff}, we carried out an additional series of runs addressing the influence of an initial wall-parallel angular deviation. We focus on the case with $(\Ga, \Bo) = (30,0.05)$ but vary the angle $-\theta_{zy}^0$ from 0 to $7.1^\circ$ (i.e., for $-\Delta_z^0$ up to $1$). The corresponding Reynolds number and aspect ratio of the bubble are $(\Rey, \upchi) = (184, 1.1)$, both comparable with that of a $1~\text{mm}$ diameter bubble rising in water. Hence, the resulting final geometry of the bubble pair at various $\Delta_z^0$ might be used to anticipate the accumulative behavior of $1~\text{mm}$ diameter bubbles in a bubble-rich near-wall layer.

Figure \ref{fig:17appen}$(a,b)$ show the obtained bubble trajectories in the wall-normal and wall-parallel planes, respectively. The paths illustrated in $(b)$ clearly indicate that, irrespective of $\theta_{zy}^0$, the TB escapes from the wake of the LB without substantially affecting the latter's trajectory. Moreover, the larger the $|\theta_{zy}^0|$, the smaller the wall-parallel separation of the two bubbles after the escape. This latter feature is more clearly identified in figure \ref{fig:17appen}$(d)$, which shows the time evolution of the wall-normal separation $\Delta_z$. In $(a)$, the paths of the LB are omitted, as only a marginal influence of $|\theta_{zy}^0|$ can be identified. Regarding the TB paths in the wall-normal plane, they start to diverge once the TB reaches the initial vertical position of the LB. Given that the pair interaction takes place in both planes simultaneously, the trend in the change of the TB's path with increasing $|\theta_{zy}^0|$ is not monotonous. Nevertheless, once the escape in the wall-parallel plane is completed, the TB begins to bounce at a fixed frequency and lateral amplitude, with values nearly identical to those of the LB (not shown). The bouncing phases of the two bubbles usually do not match, leading to periodic fluctuations in both the vertical separation and the inclination angle in the wall-normal plane, as clearly seen in figure \ref{fig:17appen}$(c,e)$.

\subsection{Influence of initial wall-normal separation and vertical displacement}
\label{sec:appA2}
In most of the results discussed in the main part of the paper, the two bubbles are initially positioned with a horizontal wall-normal separation of $X^0 = 1.5$ and a vertical displacement of $\Delta_y^0 = 8$. This appendix discusses the influence of these two geometric parameters on the paths of the two bubbles. To explore this effect, we select two typical cases discussed in \S~\ref{sec:3res}: one with $(\Ga, \Bo) = (15, 0.2)$ and the other with $(\Ga, \Bo) = (25, 0.2)$. For $(X^0, \Delta_y^0) = (1.5, 8)$, these bubble pairs exhibit a WPE (wall-parallel escape) scenario and a WNE (wall-normal escape) scenario, respectively. New results obtained with $X^0 = 1.25$ and 1.75 (while fixing $\Delta_y^0 = 8$) and with $\Delta_y^0 = 6$ and 10 (while fixing $X^0 = 1.5$) are discussed below.

\begin{figure}
    \centering
    \includegraphics[height=9.5cm]{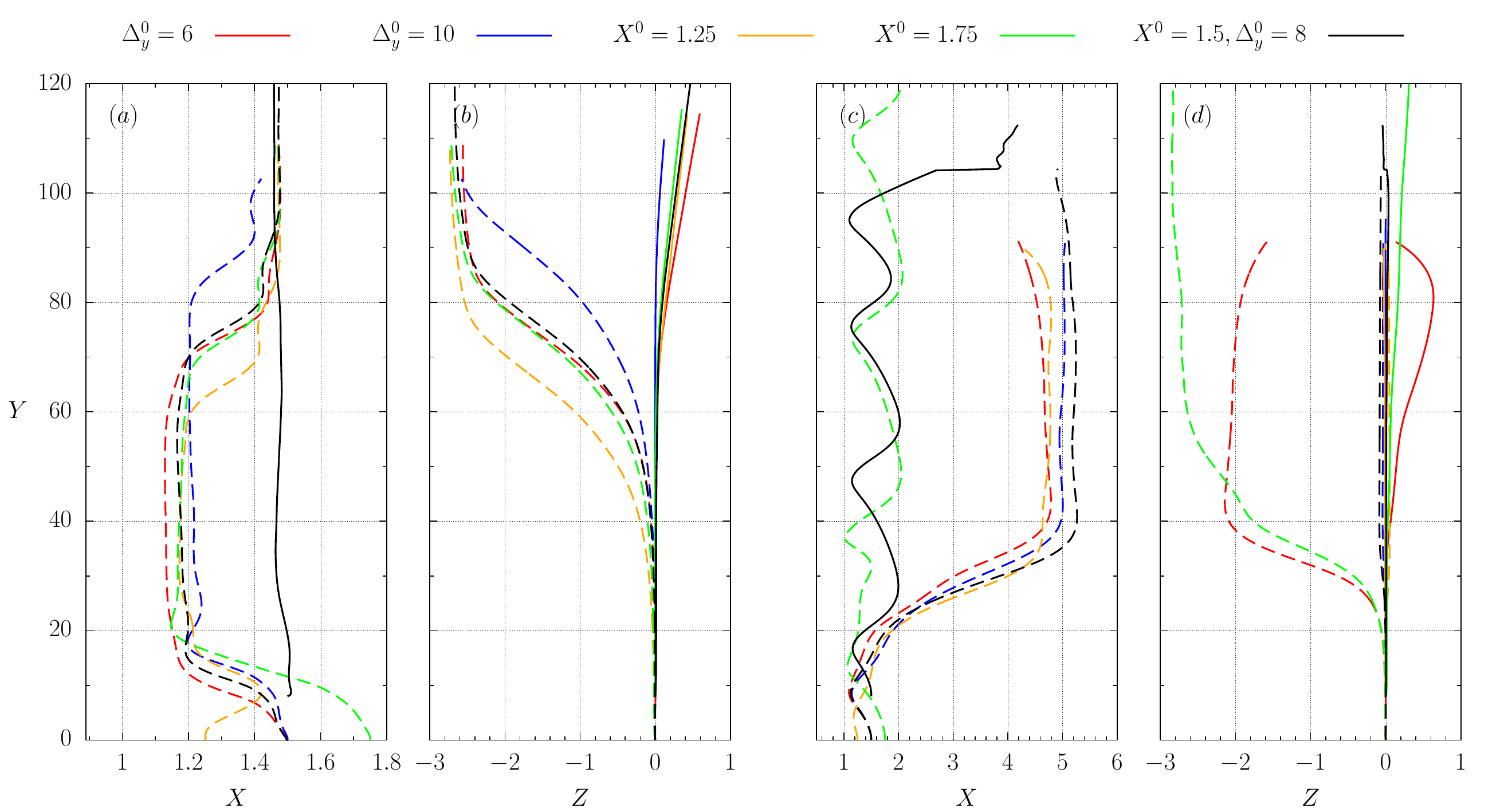}
    \caption{Influence of the initial wall-normal separation, $X^0$, and the vertical separation between the two bubbles, $\Delta_y^0$, on the paths of the LB (solid line) and the TB (dashed line) for cases with $(\Ga,\Bo) = (15,0.2)$ [panels $(a,b)$] and $(25,0.2)$ [$(c,d)$]. Panels $(a,c)$ and $(b,d)$ display the projections of the paths in the $(X,Y)$ and $(Z,Y)$ planes, respectively. In panels $(a,c)$, trajectories of the LB are shown only for $(X^0,\Delta_y^0) = (1.5, 8)$, as those at different $X^0$ and $\Delta_y^0$ exhibit similar behaviour.}
    \label{fig:18appen}
\end{figure}

Figure \ref{fig:18appen}$(a,b)$ illustrates the bubble trajectories for the case $(\Ga, \Bo) = (15, 0.2)$. These results indicate that, within the considered range of variations in $X^0$ and $\Delta_y^0$, wall-parallel escape is consistently maintained. Specifically, at a fixed initial wall-normal separation ($X^0 = 1.5$), decreasing the initial vertical separation $\Delta_y^0$ promotes the onset of wall-parallel separation, as this reduces the time required for the TB to enter the wake of the LB. Conversely, with $\Delta_y^0$ fixed at 8, reducing $X^0$ from 1.75 to 1.5 does not cause any noticeable change in the occurrence of wall-parallel separation. However, as $X^0$ further decreases to 1.25, the separation occurs earlier, starting when the vertical displacement of the TB reaches about $20R$ instead of about $40R$ as in the cases with larger $X^0$ [see dashed lines in panel $(b)$]. This observation aligns with the discussion in \S~\ref{sec:3-3-1wne_b}, where the wall was found to promote escape in the wall-normal plane. From the subtle differences between results at $X^0 = 1.5$ and $X^0 = 1.75$, it can be inferred that for a WPE scenario, this promotion no longer plays a role once the initial separation exceeds the equilibrium position achieved in the LB-only condition.

Now let us move on to the case with $(\Ga, \Bo) = (25, 0.2)$. Recall that this case lies in the unsteady flow regime where a single near-wall rising bubble follows a regular bouncing motion. In the pair configuration, the bubble path follows a WNE scenario with $(X^0,\Delta_y^0)=(1.5,8)$. Results shown in figure \ref{fig:18appen} $(c,d)$ indicate that a wall parallel separation can take place by either reducing $\Delta_y^0$ to 6 or by increasing $X^0$ to 1.75. Specifically, in the former case, the system undergoes bi-planar separation led by the TB. These drastically different behaviours compared with the case at $(\Ga, \Bo) = (15, 0.2)$ is not surprising. As already identified in section \ref{sec:3-4par_eff}, the influence of the initial geometrical configuration is difficult to be quantified when the case lies in the unsteady flow regime. In this regime, the fate of the bubble pair depends largely on the lateral position and velocity of the TB when it rises to the LB’s initial position. The difficulty lies in the fact that both bubble may undergo an wall-normal oscillation due to vortex shedding. Hence, both values do not vary monotonously with $X^0$ and $\Delta_y^0$.

To summarize, taken together with the results in \S~ \ref{sec:3-4par_eff}, it may be concluded that the regime map shown in the main part of the paper, which was obtained by setting $(X^0,\Delta_y^0)=(1.5,8)$, may still serve as a reference for different $(X^0,\Delta_y^0)$, provided that the case specified by $(\Ga, \Bo)$ lies in the steady flow regime where no vortex shedding occurs. However, the fate of the bubble pair may change drastically if the case lies in the unsteady flow regime. Providing an exhaustive regime map in the multi-parameter space defined by $(\Ga, \Bo, X^0,\Delta_y^0)$ is beyond the scope of the present work.

\section{Flow field in the wake of the LB}\label{sec:append-flowfield}

\begin{figure}[h]
    \includegraphics[height=6.5cm]{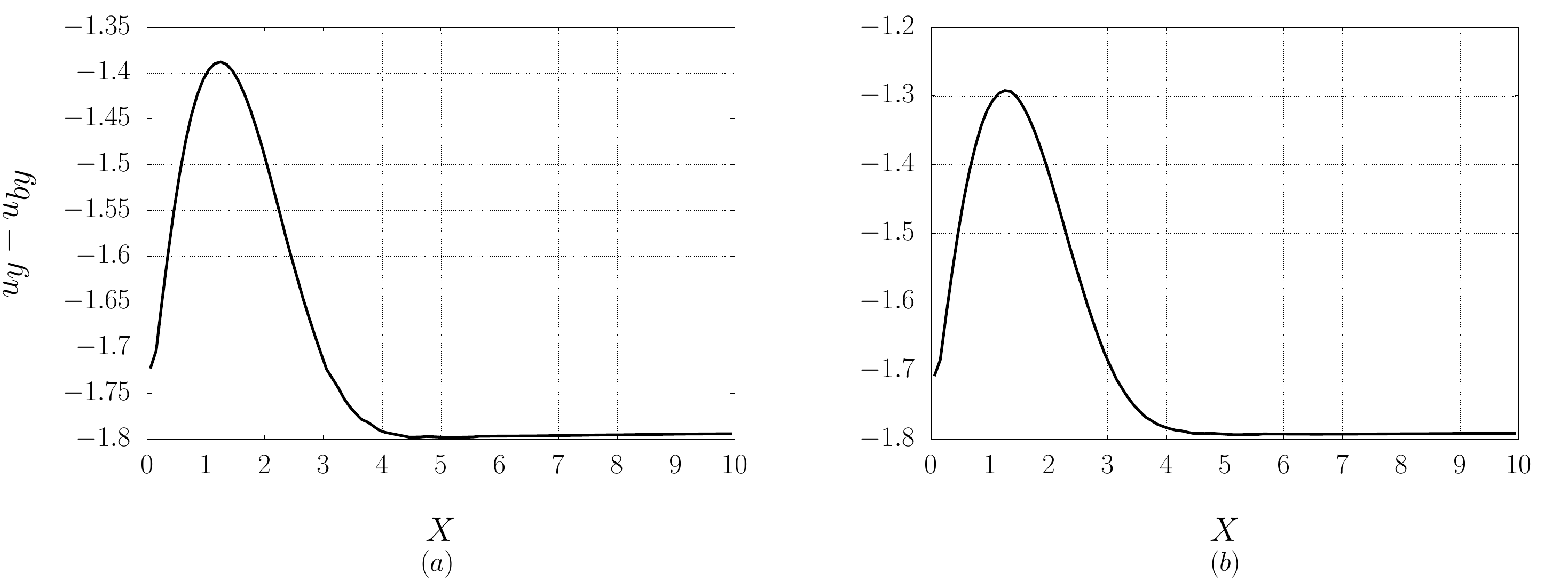}
    \caption{Variation along $X$ of the liquid rising velocity (relative to the rising speed of the LB) in the wake of the LB for the case with $(\Ga,\Bo)=(15,0.2)$ at (a) $t=7$, (b) $t=15$. The profiles are taken at a vertical distance of $6.5R$ downstream from the LB centroid.}
    \label{fig:18flowfield}
\end{figure}

As a supplement to figure \ref{fig:4wpe_r_velo}, figure \ref{fig:18flowfield} illustrates the variation along $X$ of the liquid rising speed (subtracted by the rising velocity of the LB) at a vertical distance $6.5R$ downstream of the LB for the case with $(\Ga, \Bo) = (15, 0.2)$. For the two selected time instants, it may be noted that a stronger shear flow exists on the wall-facing side, due to the no-slip condition at the wall. 
\section{Description of the movies}
\label{sec:appB}
Movie 1 - Evolution ($zy$ projection) of the isosurface $\omega_y=\pm 0.1$ of the streamwise vorticity for a bubble pair undergoing a wall-parallel escape scenario [$(\Ga,\Bo)=(20,0.2)$]. The black (respectively, light gray) thread corresponds to negative (respectively, positive) values. This movie corresponds to figures \ref{fig:5wpe_b_path} and \ref{fig:6wpe_b_velo}.

Movie 2 - Evolution ($xy$ projection) of the isosurface $\omega_y=\pm 0.1$ of the streamwise vorticity for a bubble pair undergoing a wall-normal escape scenario [$(\Ga,\Bo)=(25,0.2)$]. The black (respectively, light gray) thread corresponds to negative (respectively, positive) values. This movie corresponds to figures \ref{fig:7wne_b_path} and \ref{fig:8wne_b_geo}.

\begin{acknowledgments}
J.Z. gratefully acknowledge the supports from  National Key R $\&$ D Program of China (2023YFA1011000) and from the NSFC (12222208). P.S. acknowledges the funding of the Deutsche Forschungsgemeinschaft (DFG, German Research Foundation) through grant number 501298479.
\end{acknowledgments}
\section*{Author contributions}
H. Huang: Conceptualization, Data curation, Investigation, Visualization; 
P. Shi: Investigation, Supervision, Writing -- original draft; 
N. Elkina \& H. Schulz: Data curation, Resources, Software; 
J. Zhang: Conceptualization, Funding acquisition, Investigation, Supervision.
All authors contributed equally to Writing -- review \& editing.
\bibliography{main}

\end{document}